\documentclass[
  aps,
  prr,
  reprint,
  longbibliography,
  floatfix,
  nofootinbib,
  superscriptaddress
]{revtex4-2}

\usepackage[dvipsnames]{xcolor}
\usepackage{graphicx}
\usepackage{amsmath,amssymb,amsfonts,mathtools}
\usepackage{bm}
\usepackage[figuresright]{rotating}
\usepackage{multirow}
\usepackage{tabularx}
\usepackage{textcomp}
\usepackage{units}
\usepackage{psfrag}
\usepackage{wasysym}
\usepackage{physics}
\usepackage{soul}
\usepackage{times}
\usepackage{enumitem}
\usepackage[caption=false]{subfig}
\usepackage{tikz}

\usepackage[
  colorlinks=true,
  citecolor=blue,
  linkcolor=magenta,
  hypertexnames=false
]{hyperref}
\renewcommand{\selectlanguage}[1]{}

\newcommand{\ii}{\text{i}}
\newcommand{\e}[1]{\text{e}^{#1}}

\begin{document}

\title{Spontaneous breaking of a continuous symmetry at a non-conformal quantum critical point in one dimension}
\author{R. Flores-Calderón$^*$}
\affiliation{Technical University of Munich, TUM School of Natural Sciences, Physics Department, 85748 Garching, Germany}
\affiliation{Munich Center for Quantum Science and Technology (MCQST), Schellingstr. 4, 80799 München, Germany}
\author{M. Zündel$^*$}
\affiliation{Université Grenoble Alpes, CNRS, LPMMC, 38000 Grenoble, France}
\affiliation{Universität Innsbruck, Institut für Theoretische Physik, Technikerstraße 21a, 6020 Innsbruck, Austria}
\begin{abstract}
In this work, we present numerical evidence for the spontaneous breaking of a continuous $U(1)$ symmetry in a nearest-neighbor interacting spin-1 chain at a quantum critical point separating two XY quasi-long-range ordered phases distinguished by a spontaneously broken $\mathbb{Z}_2$ symmetry. Remarkably, the continuous symmetry breaking emerges precisely at the critical point of the discrete order parameter, suggesting a novel mechanism beyond currently established scenarios. At criticality, the XY correlations develop true long-range order, accompanied by a finite perpendicular magnetization, a zero-frequency Bragg peak in the transverse dynamical structure factor, and sharp gapless collective excitations. From complementary static and dynamical observables, we quantitatively determine the critical exponents, obtaining a dynamical exponent $z=1.50\pm0.04$ and an anomalous dimension $\eta=1.04\pm0.03$. We further show that the non-interacting continuum limit is equivalent to the recently introduced transverse quantum fluid, displaying their off-diagonal long-ranged order in one dimension. Complementing the numerical study, we derive the renormalization-group flow equations of the continuum theory to second order in the $\varepsilon$ expansion. We identify an interacting fixed point whose critical behavior differs from the Ising universality class already at two-loop order, although the perturbative exponents remain far from the numerical values.
\end{abstract}

\maketitle
\def\thefootnote{*}\footnotetext{Both authors contributed equally to this work\\
r.flores-calderon@tum.de\\
martina.zuendel@lpmmc.cnrs.fr}\def\thefootnote{\arabic{footnote}}

\section{Introduction} 

The Hohenberg-Mermin-Wagner (HMW) theorem \cite{Hohenberg1967,Mermin1966} excludes the spontaneous breaking of a continuous symmetry in systems of two or fewer spatial dimensions under specific assumptions: the system is at equilibrium, has a finite temperature and sufficiently short-ranged interactions. By contrast, it has been shown that order can be stabilized in low dimensions, if one of these conditions is not met, for instance with long-ranged interactions~\cite{Maghrebi2017,Gong2016,Romen2024,Feng2023} or in non-equilibrium systems~\cite{Vicsek1995,Toner1995,Bassler1995,Corberi2003,Sala2024,Orito2025}.

The HMW theorem is commonly derived using Bogoliubov inequalities. At zero temperature, the corresponding Bogoliubov inequality constrains the order parameter in terms of the infrared excitation energy~\cite{Takada1975,Shastry1992,watanabe23}. While Coleman's theorem~\cite{Coleman1973} excludes spontaneous breaking of continuous symmetries in one-dimensional relativistic quantum field theories, sufficiently soft infrared dispersions can evade this restriction:
a paradigmatic example of such a spontaneous symmetry breaking (SSB) in a non-relativistic system is given by the $SO(3)$-symmetric Heisenberg ferromagnet~\cite{Anderson2018,Zvonarev2007,Beekman2019,Watanabe2020}. A generalization to models that are not order-parameter conserving was given in Ref.~\cite{watanabe23} by so-called frustration-free models, whose ground states simultaneously minimize every local term in the Hamiltonian. The authors conjecture that translation-invariant, gapless, frustration-free Hamiltonians generically possess sufficiently soft infrared modes to permit the spontaneous breaking of continuous symmetries in one dimension.

Following this line of thought, an important open question is whether frustration-freeness is an essential ingredient for realizing continuous symmetry breaking in one dimension. Recently, Ref.~\cite{nahum} proposed a nearest-neighbor interacting spin-1 chain that is not frustration free. The model possesses a global $U(1)\rtimes\mathbb{Z}_2$ symmetry and, based on a coherent-spin continuum theory together with mean-field and one-loop renormalization-group analyses, was argued to exhibit spontaneous breaking of the continuous symmetry at a quantum critical point.

At the Gaussian level, the resulting continuum theory is closely related to the recently introduced transverse quantum fluid (TQF)~\cite{Radzihovsky2023} as we discuss in the end matter. TQFs constitute a new class of quasi-one-dimensional superfluids with effectively divergent compressibility, quadratic low-energy dispersion, and true off-diagonal long-range order at zero temperature. Equivalently, after a duality transformation, the Gaussian theory can be viewed as a one-dimensional realization of quantum Lifshitz physics~\cite{Bervillier2004,Grinstein1981,FradkinHuseMoessnerOganesyanSondhi2004,HsuFradkin2013}, familiar from the critical theories of two-dimensional quantum dimer models~\cite{RokhsarKivelson1988,ArdonneFradkinFendley2004,IsakovEtAl2011}. These theories share a quadratic dispersion and dynamical critical exponent $z=2$, the key ingredient underlying the softened infrared fluctuations.

In this Letter, we investigate the spin-1 model of Ref.~\cite{nahum} using a combination of non-perturbative large-scale matrix-product-state calculations and perturbative renormalization-group methods. We first introduce the lattice model and its continuum field theory. We then provide numerical evidence for XY long-range order and spontaneous breaking of the continuous $U(1)$ symmetry, and quantitatively determine the associated critical exponents from complementary static and dynamical observables. Complementing the numerical analysis, we derive the renormalization-group flow equations to second order in the $\varepsilon$-expansion. We identify the interacting fixed point, compute its critical exponents, and demonstrate that its two-loop critical behavior differs from the Ising universality class. We conclude with a discussion of the implications of our results and possible future directions.

\section{Model}

Following the model proposed in Ref.~\cite{nahum}, we will study the spin-1 Hamiltonian
\begin{align}
    H=\sum_{i=1}^L \left(-J\mathbf{S}_i \cdot \mathbf{S}_{i+1}+g_1 \left(S_i^z\right)^2+ g_2\left(S_i^zS_{i+1}^z\right)^2\right), \label{Ham}
\end{align}
where  $\mathbf{S}_i$ are the spin-1 operators at site $i$ of a periodic chain of length $L$ and we will fix through this Letter the values $J=1,\ g_2=1$. The first term is a ferromagnetic Heisenberg exchange interaction; the second and last terms represent easy-plane single-ion and double-ion anisotropy, respectively. This model has several global symmetries. First, there is an $O(2)$ symmetry which is generated by  rotations around the $z$ axis and $z$ axis spin flips. Clearly, the Heisenberg term satisfies this symmetry. The remaining terms only have a $z$-component of the spin, thus are unchanged under the $U(1)$ normal subgroup and can at most change by a sign belonging to the $\mathbb{Z}_2$ subgroup in $O(2)=U(1)\rtimes\mathbb{Z}_2 $. It is the transition from a spontaneous symmetry-broken (SSB) phase to an unbroken state of the $\mathbb{Z}_2$ subgroup that we will focus on in the following. Crucially, as we will see, the system on both sides of the transition is also in a quasi-long-range ordered (QLRO) XY phase. In the following section, we provide numerical evidence for an additional breaking of the $U(1)$ symmetry at the $\mathbb Z_2$ transition point. As has been argued in the introduction at zero temperature and for a non-relativistic theory this is neither contradicting the HMW theorem, nor Coleman's theorem.

\begin{figure}[t]
    \centering
    \includegraphics[width=0.98\columnwidth]{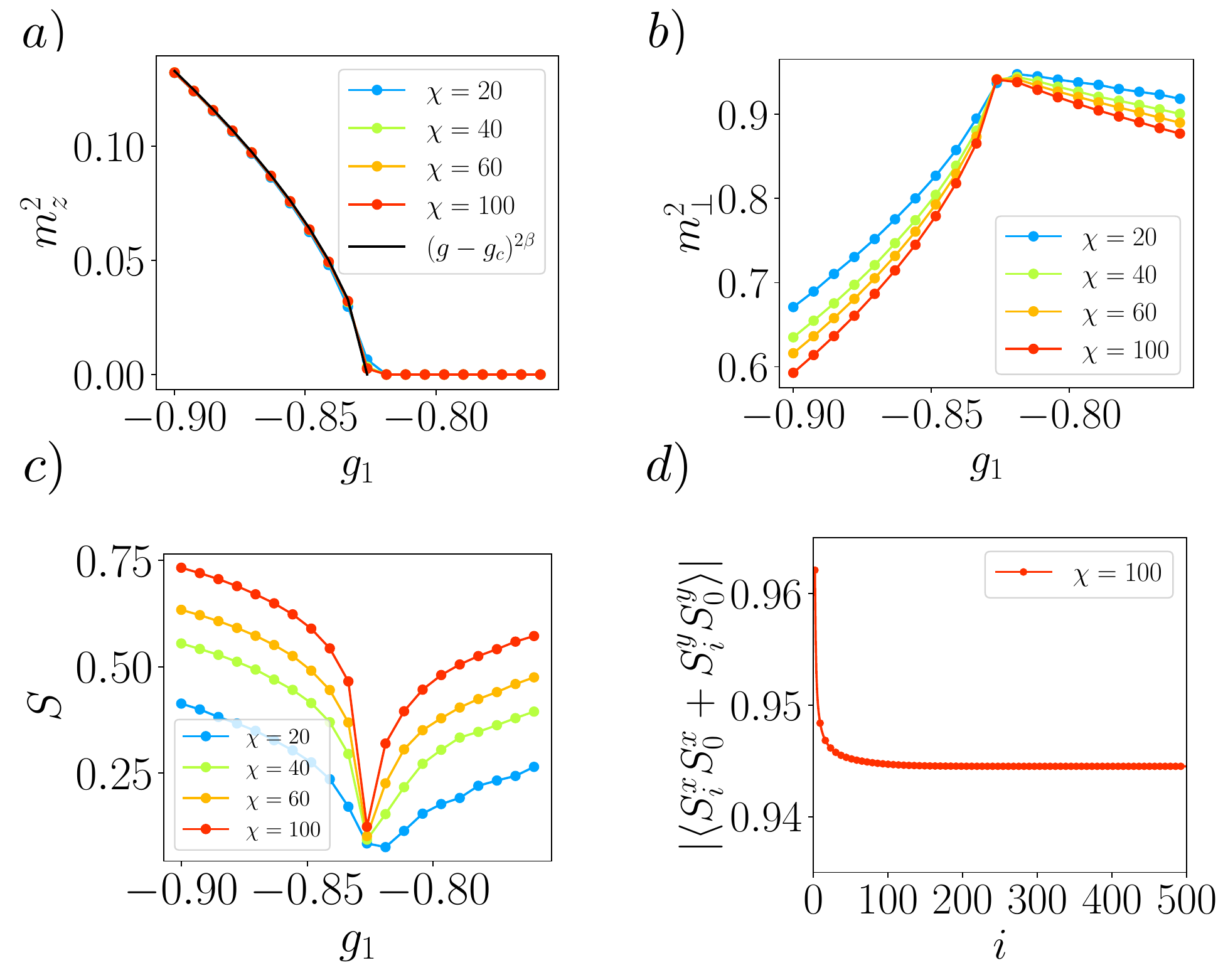}
    \caption{a) Order parameter $m_z^2$ vs coupling $g_1$ calculated in the ground state of Eq.~\eqref{Ham} using iDMRG. We vary the bond dimension $\chi$ and perform a power-law fit at $\chi=100$, finding a critical exponent of $\beta = 0.30\pm 0.01$ and the fixed point value $g_c=-0.826\pm0.005$. b) The perpendicular magnetization $m_\perp$ shows a decreasing behavior with increasing bond dimension, consistent with a QLRO of the XY type. Nevertheless, at the critical point of a) the magnetization stays approximately constant. It becomes a local maximum as a function of $g_1$.  c) Entanglement entropy as a function of $g_1$ across the critical point $g_c$ calculated with iDMRG for different bond dimensions. d) Plot of the XY correlation function for the ground-state iMPS as a function of site index $i$ for $g_1=g_c$. We observe that at large distances, the correlator behaves like a constant, i.e. the XY component is LRO.}
    \label{fig:static_obs}
\end{figure}

\begin{figure*}[ht]
    \centering
    \includegraphics[scale=0.27]{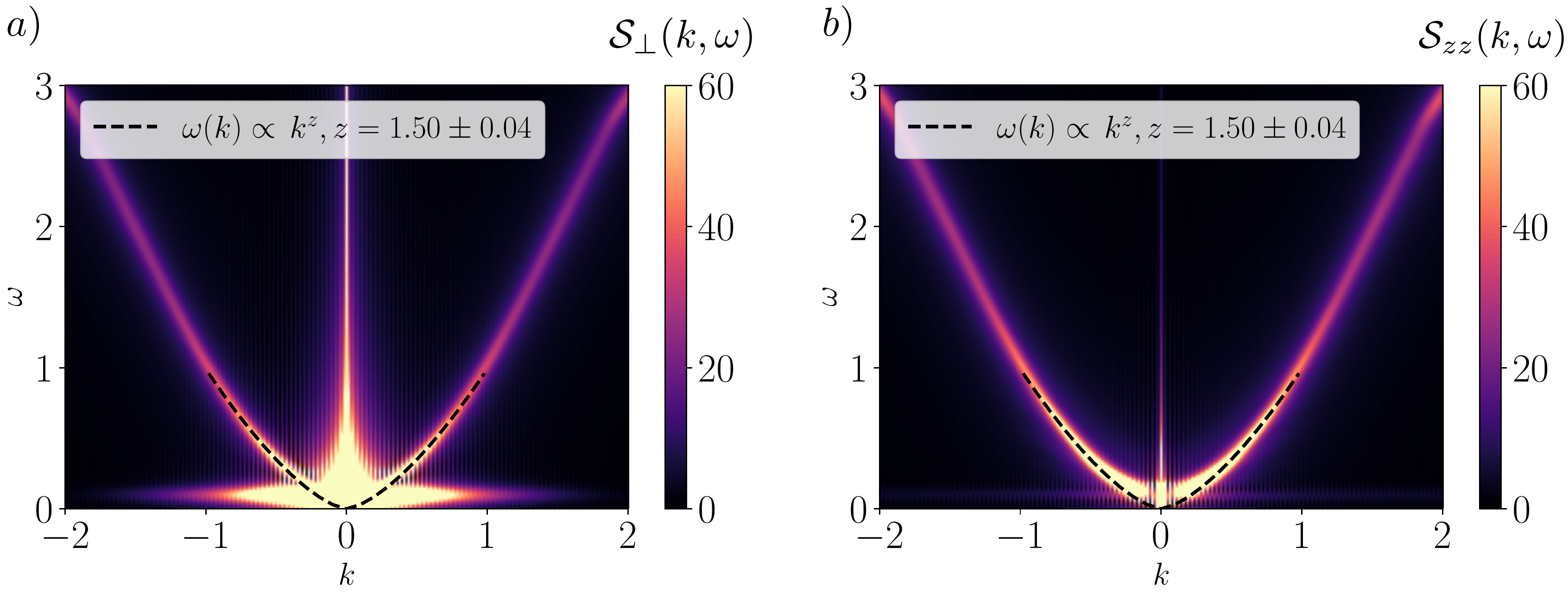}
    \caption{Dynamical spin structure factor as a function of frequency and momentum for the a) perpendicular Eq.~\eqref{DSF_perp} and b) parallel spin components Eq.~\eqref{DSF_z}. For the simulations, we use an MPS of bond dimension $\chi=150$, extent $L=400$, the time evolution is done for $\Delta t =0.1$ and 1200 time steps. A Gaussian window function with $\sigma=0.2$ is applied to the time data. We extract the maxima of $\mathcal{S}_{zz}$ and perform a log-log fit of the data as a function of frequency and momentum for long-wavelengths, shown in the inset of b). We find the relation $\omega\propto k^z$ with $z=1.50\pm 0.04$.}
    \label{fig:Spectral_functions}
\end{figure*}
 \section{Static observables}
 We use the the infinite density matrix renormalization group (iDMRG) algorithm \cite{white1992dmrg,white1993}, performed with the TeNPy library \cite{tenpy2024,hauschild_efficient_2018}, to first calculate the ground state of Eq.~\eqref{Ham}, and then, using the time-evolving block decimation (TEBD) algorithm \cite{Vidal2003,Vidal2004,White2004} we evolve the state in time. In Sec.~S1 of the supplementary material (SM), we give more details on the tensor network methods and on working directly in the thermodynamic limit by using the infinite matrix product state (iMPS). Let us look first at the $z$-magnetization density in the ground state, defined as $m_z=\tfrac{1}{L}\sum_i \expval{S_i^z}$. We perform the entanglement-scaling analysis in the SM, similar analysis were performed in Refs.~\cite{Hastings2007,Vidal2007,ScalingTagliacozzo,CumulantsiDMRG,stojevic_conformal_2015}. We compute this for increasing bond dimension $\chi$ as a function of the parameter $g_1$ as shown in Fig.~\ref{fig:static_obs}a. For parameter values $g_1$ less than approximately $-0.82$ we find $m_z\neq 0$ consistent with a $\mathbb{Z}_2$ SSB phase. Additionally, the magnetization has very similar values for different bond dimensions. In contrast, for $g_1$ bigger than $-0.82$, the magnetization has a very small value close to zero $m_z$, consistent with an unbroken $\mathbb{Z}_2$ phase. We are interested in the value of the tuning parameter $g_1$ for which the magnitude of the magnetization first goes to zero. The value of the coupling $g_c$ can be deduced by performing a fit of the order parameter $m_z$ to a power-law. We find for the bond dimension $\chi=100$ a value $g_c=-0.826\pm0.005$. The behavior of $m_z$ for $g_1<g_c$ is fitted to the critical exponent $\beta$ by
\begin{align}
    m_z^2\propto \abs{g_1-g_c}^{2\beta}\;,\qquad\beta={0.30\pm0.01}\;,\label{m_z}
\end{align}

As an additional consistency check of the critical scaling, we analyze the equal-time connected longitudinal correlator defined as $\expval{S_i^zS_0^z}_c=\expval{S_i^zS_0^z}-\expval{S_i^z}\expval{S_0^z} $. The full bond-dimension dependence and fitting procedure are given in the SM, Sec.~S2. At the critical point, we find
\begin{equation}
    \expval{S_x^zS_0^z}_c\propto \dfrac{1}{\abs{x}^{2\Delta_\psi}}\;, \qquad \Delta_\psi= 0.76 \pm 0.03\;, \label{corr_z}
\end{equation}
Here $\Delta_\psi$ is the scaling dimension of the lattice operator $S^z$, represented by the field $\psi$ in the continuum theory. Using the quantum-critical scaling relation $
\Delta_\psi= \frac{d+z+\eta-2}{2}$,
with $d=1$ the spatial dimension, $z$ the dynamical critical exponent, and $\eta$ the anomalous dimension. The mean field exponents of Eq.~\eqref{eq:L0_D} are given by $z=2,\,\eta=0$. We expect them to change as we renormalize the theory. Indeed, this is in the 2-loop RG section. Here, we deduce the numerical relation $z+\eta = 2.54 \pm0.03$ consistent with the independent dynamical estimates discussed below.

To diagnose the $U(1)$ spontaneous symmetry breaking of the model at the critical point we measure the value of the perpendicular magnetization density squared $ m_\perp^2=\tfrac{1}{L}\sum_i \expval{S_i^x}^2+\expval{S_i^y}^2$. The order parameter is the one-point function $m_\perp$ shown in Fig.~\ref{fig:static_obs}b as a function of $g_1$: If it is non-zero in the thermodynamical limit, it implies spontaneous breaking of the $U(1)$ symmetry. It is worth mentioning that the iDMRG method works directly in the thermodynamic limit, symmetry breaking is thus allowed for our iMPS simulations. We observe that for both sides of the transition $m_\perp\neq 0$, but as the bond dimension is increased, the magnetization decreases. In both cases there is an increase in the entanglement entropy and the correlation length, see Sec.~S2 in the SM. Nevertheless, at $g_c$ we have the slowest growth. 
These results are consistent with the ground states away from the phase transition having no true LRO, only QLRO, and $m_\perp\rightarrow 0$ in the thermodynamic limit.

For the critical point $g_1=g_c$ the situation is different. As the bond dimension is increased, the squared magnetization $m_\perp^2$ converges to a non-zero value. Furthermore, the plateau value of the XY correlations $\expval{S_i^xS_0^x+S_i^yS_0^y}$ in Fig.~\ref{fig:static_obs}d and the magnetization at the critical point in Fig.~\ref{fig:static_obs}b coincide.
Let us comment here on what these numerical results imply regarding LRO and $U(1)$ SSB. Within our iMPS ansatz, which assumes translational invariance and up to the bond dimensions we reach, a converged non-zero value of $m_\perp^2= \expval{S_i^x}^2+\expval{S_i^y}^2$ implies $U(1)$ SSB. We give the explicit proof in Sec.~S3 for an iMPS. Additionally, a plateau in the value of the XY correlations, in contrast to the logarithmic decay away from $g_1=g_c$, indicates LRO up to the maximum separation presented in Fig.~\ref{fig:static_obs}d.
 
It is worth noting that the order parameter $m_z$ goes to zero at the same point that the XY order parameter $m_\perp$ has a maximum. This is consistent with the intuition that the increased fluctuations of the $z$-component of the spin close to the critical point imply reduced fluctuations for the perpendicular components due to the fixed spin-length constraint and the spin commutation relations. 

\section{Dynamical structure factor}

To obtain the complete set of critical exponents and further confirm the XY LRO, we must analyze the dynamics of the system. We calculate two dynamical structure factors
\begin{align}
    & \mathcal{S}_{zz}(k,\omega) = \sum_{j}\!\int\! \dd t\,\e{\ii(\omega t-kj)}\expval{S^z_j(t)S^z_0(0)}, \label{DSF_z} \\
    & \mathcal{S}_{\perp}(k,\omega) = \sum_{j}\!\int\! \dd t\,\e{\ii(\omega t-kj)}\expval{S^x_j(t)S^x_0(0)+S^y_j(t)S^y_0(0)}. \label{DSF_perp}
\end{align}
The first quantity details the dynamics of the $z$-component of the spin, while the second of the perpendicular components; here we used the lattice spacing $a=1$. We are interested in the universal regime of the low-frequency and long-wavelength limit. This regime contains the critical properties of the quantum critical point. We plot the results at the critical point $g_1=g_c$ in Fig.~\ref{fig:Spectral_functions} for a system of $L=400$ sites, $\chi_{\max}=150$ with 1200 time-steps each with $\Delta t = 0.1$. We observe sharp modes in the spectrum of both $\mathcal{S}_{zz}$ and $\mathcal{S}_{\perp}$. These modes come down in energy and touch the $\omega=0$ point at $k=0$, consistent with a gapless critical point.
The spectral weight has a strong delta-like peak in $\mathcal{S}_{\perp}$ as seen in Fig.~\ref{fig:Spectral_functions}b, this peak is the Bragg peak consistent with $U(1)$ SSB. We further analyze the critical properties of the dynamical structure factor by extracting the maxima of the $\mathcal{S}_{zz}(k,\omega)$ spectral values for small $k$ and fitting the data to a power-law $\omega\propto k^z$, as seen in the inset of Fig.~\ref{fig:Spectral_functions}b. The result is a dynamical critical exponent of 
\begin{equation}
    z = 1.50\pm0.04,\quad  \eta = 1.04\pm0.03, 
\end{equation}
where the previous numerical relation determines the anomalous dimension. We explain details on the fits and error estimates in the SM. Remarkably, the numerical value of the anomalous dimension, together with the dynamical critical exponent, is consistent with the first renormalization condition found in the effective field theory. Note in particular that the approximate value of $z$ is also found for the perpendicular dynamical structure factor. A further validation of the exponent $\eta$ is obtained by fitting the correlator in momentum-space $S_{zz}(k,\omega=0)$, which is expected to scale as $k^{-2+\eta}$. The resulting value agrees within the errorbars with that extracted from the static correlators, as is shown in the SM. It is worth mentioning that from our numerical calculations, the streak in $S_{zz}(k=0,\omega)$ and $S_{\perp}(k=0,\omega)$ doesn't seem to be an artifact. Instead, a power-law decay is observed $S_{zz}(k=0,\omega)\propto 1/\omega^m$ with $m=1.1\pm 0.1$ as detailed in Sec.~S4 of the SM.

\section{Field-theoretic prediction}

In the continuum limit, under the assumption of a small order parameter $\psi$ close to the transition and keeping only RG-marginal and relevant terms, see SM, we map the partition function with the  Hamiltonian of Eq.~\eqref{Ham} to a theory with the following Euclidean Lagrangian 
(in imaginary time $\tau$)
\begin{align}
    \mathcal L_{1} = \ii \psi\partial_{\tau}\theta + \tfrac J 2 \big[ (\partial_x\psi)^2+(\partial_x\theta)^2\big]
    +\tfrac{g_1+g_2}{2}\,\psi^2+\tfrac{g_2}{4}\,\psi^4\;,
\end{align}
where $\theta(x,\tau)\in[0,2\pi)$ is the azimuthal angle field of the spin in the coherent state basis and $\psi(x,\tau)$ is the field corresponding to the $z$-component of the spin. The first term in Eq.~\eqref{eq:L0_D} is the Berry phase term, present in any spin system. The imaginary time is defined in the interval $\tau \in (0,\beta]$, with $\beta=1/T$, $T$ the temperature, which we take to be zero in the following. The origin of this term relies on the quantization of the spin, here $S=1$. The Berry phase term is of topological origin and allows the theory to differ from usual $O(2)$ statistical models. Importantly, the coefficient of the term will not renormalize, and the RG equations will depend strongly on this. Additionally, the first term tells us that $\theta$ and $\psi$ are conjugate variables. Furthermore, the $\theta$ field is gapless while the $\psi$ field has a mass $\tfrac 12(g_1+g_2)$ and a $\psi^4$ interaction parametrized by the coupling $g_2/4$.

\section{Dimensional regularization and Ward identities}
We perform a perturbative renormalization and a simultaneous expansion about the upper critical dimension $d_c=2$. To simplify the calculation, we start from the massless Lagrangian
\begin{equation}
\label{eq:L0_D}
\mathcal{L}[\theta,\psi]
= \mathrm i\,\psi\, \partial_\tau \theta
+ \tfrac{K}{2}\,(\nabla\theta)^2
+ \tfrac{1}{2}\, (\nabla\psi)^2
+ \tfrac{g}{4!}\,\psi^4
\;, 
\end{equation}
and introduce the mass as a perturbation in a second step. We introduce in the following the renormalized Lagrangian $\mathcal L_R$ by multiplicative renormalization. The bare couplings are related by renormalization constants $Z$ to the renormalized couplings, indicated with an index $R$:
\begin{equation}
\theta = Z_\theta^{1/2}\, \theta_R\,,\;
\psi  = Z_\psi^{1/2}\, \psi_R\,, \;
K   = \frac{Z_{K}}{Z_{\theta}} K_R \,,\;
g   = \frac{\mu^{\varepsilon}Z_{g}}{Z_{\psi}^2} g_R\,,
\end{equation}
where the scale $\mu$ is the renormalization scale, $g_R$ is a dimensionless coupling. The effective action $\Gamma_R$, generator of the one-particle irreducible correlation functions, is related by renormalization to the Legendre transformation $\Gamma[\phi]=J_a\phi_a-W[J]$ of the Schwinger functional $\e{W[{J}]}=\int\mathcal D[{\varphi}]\,\mathrm e^{-S[{\varphi}]+J_a\varphi_a}$, here field $\varphi$, mean field $\phi$ and sources $J$ have two components: $\psi$ and $\theta$. We use de Witt notation~\cite{DeWitt1965} and Einstein's sum convention. We abbreviate functional derivatives as $S^{\varphi_1\dots\varphi_n}=\frac{\delta^n}{\delta\varphi_1\dots\delta\varphi_n}S$.

We deduce two renormalization conditions constraining the $Z$-factors: First, the nonrenormalization of the Berry phase term $\partial_\omega\Gamma^{\psi\theta}_{R}\big|_{\text{NP}}=1$ at the normalization point (NP) leading to
$Z_\psi =Z_\theta^{-1} \equiv Z$, meaning the conjugate fields $\theta_R$ and $\psi_R$ are renormalized by one constant only. Second, $\Gamma^{\theta\theta}=S^{\theta\theta}$ leading to $Z\,K_R=K$ and $Z_K=1$. This Ward identity is equivalent to the observation that the self-energy $\Sigma^{\theta\theta}=0$ is identical to zero to all loop orders since there are no bare vertices that have $\theta$-legs. In the SM, we explain how the Ward identities can be derived from the infinitesimal shift $\theta(x,t)\mapsto\theta(x,t)+\epsilon(x,t)$. 
The regularized Lagrangian, therefore, reads
\begin{align}
\label{eq:SR_D2_main}
\mathcal L_R
=
\ii\psi_R\partial_\tau \theta_R
+ \tfrac{K_R}{2}  (\nabla\theta_R)^2
+ \tfrac{1}{2}Z (\nabla\psi_R)^2%\nonumber\\&
+ \tfrac{g_R}{4!}\mu^{\varepsilon}Z_g \psi_R^4
\,.
\end{align}
Astonishingly, the number of independent renormalization constants are those of the textbook~\cite{ZinnJustin_book} massless $\phi^4$-theory.
Since we are interested in the behavior of the correlation functions near criticality by tuning the relevant parameter, i.e. the mass, we add a source term for the monomial $\psi^2_R$ to the action. We write
\begin{align}
    \mathcal L_R[\theta_R,\psi_R,\psi_R^2] = \mathcal L_R[\theta_R,\psi_R]+\frac{1}{2} m_R^2 Z_2 \psi_R^2\;,
\end{align}
where with respect to the initial action $m^2 = m_R^2\,Z_2\,Z^{-1}$.
In the SM, we detail how to carry out the calculation in dimension $d=2-\varepsilon$ in the minimal subtraction $\overline{MS}$ scheme. In the following, we use an effective coupling, absorbing the loop factors $\tilde g \, \coloneqq\, \frac{\sqrt{K_R}}{8\pi} g_R$.

At one-loop order, the flow equations are equivalent to the Ising ones, where the expansion is done around the upper critical dimension $d_c^{\text{(Ising)}}=4$, instead of $d_c=2$. Therefore, the critical exponents also correspond to the Ising universality class, in agreement with Ref.~\cite{nahum}.

\section{Two-loop results}

In a very similar way to the two-loop calculation for the Ising model~\cite{ZinnJustin_book}, we calculate the corrections to two-loop order; all details are given in the SM. Here, we show the resulting flow equations, which differ from the Ising ones
\begin{align}
    \tilde \beta (\tilde g) &= -\varepsilon \tilde g + \frac 32 \tilde g^2+\gamma_1\tilde g^3 + \mathcal O(\tilde g^4)\;,\\
    \eta(\tilde g) &= \frac{1}{27} \tilde g^2\,+\,\mathcal O(\tilde g^3)\;,\\
    \eta_2(\tilde g) &= -\frac{\tilde g} 2 + \gamma_2 \tilde g^2+ \mathcal O (\tilde g^3)\;,
\end{align}
with $\gamma_1\coloneqq-\frac{17}{18} -3\ln 2+\frac 32 \ln 3$ and $\gamma_2\coloneqq  \frac12\ln 2 - \frac 14\ln 3 + \frac 7{54}$. The positive, non-zero fixed point solution is given to second order in epsilon by $\tilde g^*=\frac 23 \varepsilon-\frac 8 {27}\gamma_1\varepsilon^2+\mathcal O(\varepsilon^3)$. Below the upper critical dimension, the interacting fixed point has the structure of an ordinary critical point, analogous to the Wilson--Fisher fixed point~\cite{Goldenfeld1992}. We obtain first corrections to the mean field value of the critical exponents $\eta = \frac{4}{243}\varepsilon^2$, hence $z=2-\frac{2}{243}\varepsilon^2$ and $2\nu = 1+\tfrac 16 \varepsilon + [\tfrac{1}{36} -\tfrac 29\gamma_2-\tfrac{2}{27}\gamma_1]\varepsilon^2$. At $\varepsilon=1$ this gives 
\begin{align}
    \eta\simeq0.016\,,\;z\simeq1.992\,,\;\nu\simeq0.626\;,
\end{align}
compared to the one-loop (Ising-like) values $\eta=0$, $z=2$, and $\nu\simeq0.583$.

At $\varepsilon=1$, where the expansion is strictly speaking no longer controlled, hence the perturbative approach can only be validated \emph{a posteriori}. The fixed-point value $\tilde g^*\simeq 1.07$ changes appreciably compared to the one-loop result. The critical exponents, however, are less strongly affected. This situation is reminiscent of the $\varepsilon$-expansion of the three-dimensional Ising universality class: at low orders the fixed point itself receives sizeable corrections and a reliable estimate of the anomalous dimension $\eta$ requires higher orders together with a dedicated resummation procedure, typically based on a Borel--Leroy transformation combined with a conformal mapping~\cite{LeGuillouZinnJustin1980,GuidaZinnJustin1998,KompanietsPanzer2017, Schnetz2023}. By contrast, in the Ising model the exponent $\nu$ is comparatively close to its accepted value already at two-loop order. This is not the case for the present calculation: the deviations from the numerical results are sizeable for all exponents. It indicates that the present expansion can at best approach them slowly. Higher orders in perturbative RG would be needed to assess the convergence of the expansion and to obtain a quantitative determination of the exponents.

\section{Discussion and outlook} 
In this work, we investigated the quantum phase transition in a spin-1 chain between two quasi-long-range-ordered XY phases distinguished by the onset of a nonzero $\mathbb{Z}_2$ order parameter, combining tensor-network simulations with field-theoretical arguments. Our numerical results show that the transition described by Eq.~\eqref{Ham} is continuous and characterized by the Ising order parameter $m_z$. At criticality, we find robust numerical evidence for long-range transverse order, including a finite perpendicular magnetization $m_\perp$, non-decaying XY correlations in the thermodynamic limit, and a zero-frequency, zero-momentum Bragg peak in the transverse dynamical structure factor. These results support the interpretation that the continuous $U(1)$ symmetry is spontaneously broken at the $\mathbb{Z}_2$ critical point. We further analyze the intensity at $k=0$ in the longitudinal and transverse dynamical structure factors, finding behavior consistent with intrinsic critical fluctuations. From the combined static and dynamical scaling analyses, we obtain the critical exponents $\eta = 1.04\pm0.03$, $z = 1.50\pm0.04$, and $\beta = 0.30\pm0.01$.

The value of the dynamical exponent is remarkably close to the one-dimensional KPZ value $z=3/2$. The KPZ universality class governs the scaling and statistics of countless non-equilibrium phenomena in nature, and it was also found in quenched quantum spin chains~\cite{Ljubotina2019,Wei2022,Ye2022,Rosenberg2024}. However, to the best of our knowledge, this is the first time $z\simeq3/2$ was found in a quantum system at equilibrium: It is striking, especially, since the KPZ equation itself is a stochastic evolution equation that is out-of-equilibrium. Only in one dimension, it fulfills a time-reversal symmetry.

On the analytical side, complementary to the numerical study, we performed a perturbative RG calculation using dimensional regularization. We derived the full set of flow equations to second order in the coupling. These flow equations differ from those of the Ising model, and the resulting critical exponents at the interacting fixed point also deviate from the Ising universality class at second order in $\varepsilon$. In particular, we obtained a finite anomalous dimension $\eta$ and a dynamical critical exponent $z$ that differs from its bare value, in contrast to the one-loop result of Ref.~\cite{nahum}.

The perturbative corrections to the mean-field exponents, however, remain rather small even at two-loop order. Since the exponents extracted numerically in one dimension differ substantially from their mean-field values, this suggests that the present $\varepsilon$-expansion can at best approach the microscopic critical behavior slowly. Moreover, since the actual exponents are so far from the Gaussian fixed point, the assumption of retaining the operators that are relevant or marginal by bare power counting underlying the derivation of the Lagrangian~\eqref{eq:L0_D} might not be valid. As discussed in the SM, a more general Landau-Ginzburg-type Lagrangian contains additional interactions that lift the second Ward identity and may therefore modify the critical behavior.

In any case, the discrepancy indicates that a quantitative description of the transition might require going beyond the perturbative expansion around the Gaussian fixed point. Among these methods are tensor networks (this work), Monte Carlo and functional RG. The latter could give further insight also about the correct effective field-theoretical description at and near the critical point and help to discriminate between the two suggested microscopic descriptions of the transition of the underlying spin model.

More broadly, our results establish a microscopic one-dimensional quantum spin model exhibiting an unconventional critical point outside the standard conformal paradigm. The simultaneous appearance of $\mathbb{Z}_2$ criticality and signatures of continuous $U(1)$ symmetry breaking raises the intriguing possibility that the criticality of one order parameter can induce long-range order associated with another. Establishing the mechanism behind this phenomenon could open a new avenue for understanding and engineering unconventional ordered phases in low-dimensional quantum systems. Another important direction is to determine whether the observed KPZ-like dynamical critical exponent can ultimately be accounted for within the present continuum theory, despite the apparent slow convergence of the $\varepsilon$ expansion, or whether additional interactions beyond those retained in the present continuum theory become relevant at criticality. Other intriguing directions include mapping out the full phase diagram as a function of $(g_1,g_2)$, particularly the vicinity of the $SU(2)$-symmetric point at the origin, and exploring experimental realizations in platforms such as Rydberg atom arrays~\cite{Sompet2022,Luo2024}, quantum gas microscopes~\cite{Moegerle2025}, and possibly magnonic systems~\cite{Chauhan2020}.

\textit{Acknowledgements}.-- We gratefully acknowledge fruitful discussions with M. Heinrich, A. Al-Eryani, L. Gosteva, S. Ray, J. Moore, M. Buchhold, N. Dupuis and F. Pollmann. MZ thanks RFC for bringing up Ref.~\cite{nahum}. This research was undertaken in part at the  MPI-PKS in Dresden, Germany.

\appendix

\section{Connection to transverse quantum fluids}
At the Gaussian level, the continuum theory is described by the conjugate fields $\theta$ and $\psi$ with Euclidean action
\begin{align}
S_0
=
\int d\tau dx\,
\left[
\ii\psi\,\partial_\tau\theta
+
\frac{\rho}{2}(\partial_x\theta)^2
+
\frac{\kappa}{2}(\partial_x\psi)^2
\right]\;.
\end{align}
The important feature of this action is the absence of a local density-cost term proportional to $\psi^2$. Integrating out the conjugate field $\psi$ gives the effective quadratic action
\begin{align}
S_{\mathrm{eff}}[\theta]
=
\frac12
\int_{\omega,k}
\left[
\rho k^2
+
\frac{\omega^2}{\kappa k^2}
\right]
|\theta(\omega,k)|^2\;.
\end{align}
The equal-time phase fluctuation is therefore infrared finite,
\begin{align}
\left\langle |\theta(k)|^2\right\rangle_{\tau=0}
=
\int\frac{d\omega}{2\pi}
\frac{\kappa k^2}{\omega^2+\rho\kappa k^4}
=
\frac12\sqrt{\frac{\kappa}{\rho}}\;.
\end{align}
Consequently,
\begin{align}
\left\langle [\theta(x,0)-\theta(0,0)]^2\right\rangle
&=
\sqrt{\frac{\kappa}{\rho}}
\left[
\frac{\Lambda}{\pi}
-
\frac{\sin(\Lambda x)}{\pi x}
\right]\;,
\end{align}
where $\Lambda$ is an ultraviolet cutoff. Thus, the phase contribution to the single-particle density matrix approaches a nonzero constant,
\begin{align}
\nonumber
\left\langle e^{i\theta(x,0)}e^{-i\theta(0,0)}\right\rangle
&=
\exp\left[
-\frac12
\left\langle [\theta(x,0)-\theta(0,0)]^2\right\rangle
\right] \\
&
\xrightarrow{|x|\to\infty}
\exp\left[
-\frac{\Lambda}{2\pi}
\sqrt{\frac{\kappa}{\rho}}
\right]
\neq 0\;.
\end{align}
The limiting value is nonuniversal because it depends on the ultraviolet cutoff, but its finiteness shows that the noncompact Gaussian theory has true off-diagonal long-range order rather than algebraic quasi-long-range order, provided that compact topological defects do not proliferate.

This infrared mechanism is closely related to the transverse quantum fluid (TQF) theories of Ref.~\cite{Kuklov2024}. Identifying
\begin{align}
\theta \leftrightarrow \phi,\qquad
\psi \leftrightarrow n,\qquad
\rho \leftrightarrow n_s,\qquad
\kappa \leftrightarrow \chi\;,
\end{align}
the Gaussian action has the same structure as the coherent TQF action. In both cases, the absence of a local $n^2$ or $\psi^2$ term makes the effective phase action nonlocal in space, producing the term $\omega^2/(\kappa k^2)$ instead of the usual Luttinger-liquid form. As a result, the equal-time phase fluctuations remain finite in the infrared. This should be contrasted with an ordinary Luttinger liquid, where a local density term corresponding to a finite compressibility gives
\begin{align}
\left\langle |\theta(k)|^2\right\rangle_{\tau=0}
\sim \frac{1}{|k|}\;,
\end{align}
leading to logarithmically divergent phase fluctuations and therefore only algebraic quasi-long-range order.

More generally, the TQF is one realization of a broader mechanism by which nonlocal couplings suppress the infrared fluctuations that preclude long-range order in an ordinary one-dimensional Luttinger liquid. Long-range hopping can suppress phase fluctuations and stabilize $U(1)$ order~\cite{MaghrebiGongGorshkov2017PRL,FrerotNaldesiRoscilde2017PRB,Orignac2025PRB,GuptaProkofevPupillo2026PRL}. Long-range density interactions instead suppress fluctuations of the dual displacement field: Coulomb interactions produce quasi-Wigner correlations, whereas interactions that decay more slowly than Coulomb can yield a true Wigner crystal with broken translation symmetry~\cite{Schulz1993PRL,DavietDupuis2020PRL}. Coupling a one-dimensional mode to gapless reservoir degrees of freedom provides a temporally nonlocal route to long-range order~\cite{WeberLuitzAssaad2022PRL,RibeiroMcClartyRibeiroWeber2024PRB}; in the incoherent TQF, integrating out transverse Luttinger liquids generates a $|\omega|$ kernel and superfluid long-range order~\cite{Kuklov2024}.

The gapless structure of the QFT is protected by the geometric origin of the conjugate field, for instance as a displacement mode, whereas in the continuum theory considered in the main text it appears at the critical point where the local $\psi^2$ term vanishes. Away from this point, the theory crosses over to an ordinary Luttinger liquid, and the Gaussian off-diagonal long-range order is replaced by algebraic quasi-long-range order. The main text, as opposed to this appendix, considers the interacting theory, where it remains open whether a similar mechanism is present at the interacting critical point.

\section{Brief explanation of tensor network methods}
\label{sec:tensor_network_methods}
We use matrix-product-state (MPS) methods to study the ground state and dynamical properties of the spin-1 model Eq.~\eqref{Ham}. Here we dsicuss only the minimal important aspects of the method, for a deeper review see Refs. \cite{white1992dmrg,white1993,tenpy2024,hauschild_efficient_2018}. An MPS many-body wavefunction can be written as
\begin{equation}
    \ket{\psi} =
    \sum_{\{s_i\}}
    A_1^{s_1} A_2^{s_2}\cdots A_L^{s_L}
    \ket{s_1s_2\cdots s_L},
\end{equation}
where $s_i=-1,0,1$ is the local spin-1 index in site $i$.  We introduced the  matrices $(A_i^{s_i})_{\alpha\beta}$ with internal dimension (auxiliary index) $\alpha = 1,\dots,\chi_{i} $ and $\beta = 1,\dots,\chi_{i+1}$. The matrices are contracted over auxiliary indices whose maximum dimension $\chi_i$ is the bond dimension $\chi=\text{max}( \{\chi_i\})$.  The bond dimension is the maximum Schmidt rank retained across a cut, and therefore bounds the half-chain entanglement entropy as
\begin{equation}
    S_{\mathrm{vN}}\leq \ln \chi .
\end{equation}
Finite bond dimension $\chi$ is the basic approximation in all tensor-network calculations below. For static quantities we mostly use infinite-system DMRG (iDMRG), which is a variational optimization for a periodically repeated or infinite MPS (iMPS) unit cell directly in the thermodynamic limit, $A=A_i, \forall i$ and $\chi_i=\chi$. We can write an iMPS as 
\begin{equation}
    \ket{\psi} =
    \sum_{\{s_i\}}
    \cdots A^{s_{n-1}}A^{s_n} A^{s_{n+1}}\cdots 
    \ket{\cdots s_{n-1}, s_n, s_{n+1}\cdots},
\end{equation}
The paramagnetic product state for spin-$1/2$ given by $\ket{\cdots \leftarrow \leftarrow \cdots}$ is a trivial iMPS example with bond dimension $\chi=1$ and $A^{\uparrow}=1/\sqrt{2}$ and $A^{\downarrow}=-1/\sqrt{2}$.
Thus iDMRG correlation functions are not restricted by a finite length. The limitation is instead finite entanglement.  At a critical point the exact correlation length diverges, while the MPS correlation length is cut off at a bond-dimension-dependent value $\xi_\chi$.  Static exponents must therefore be extracted from windows that are longer than the microscopic lattice scale but shorter than the finite-entanglement cutoff.

\begin{figure}[t]
    \centering
    \includegraphics[width=0.9\columnwidth]{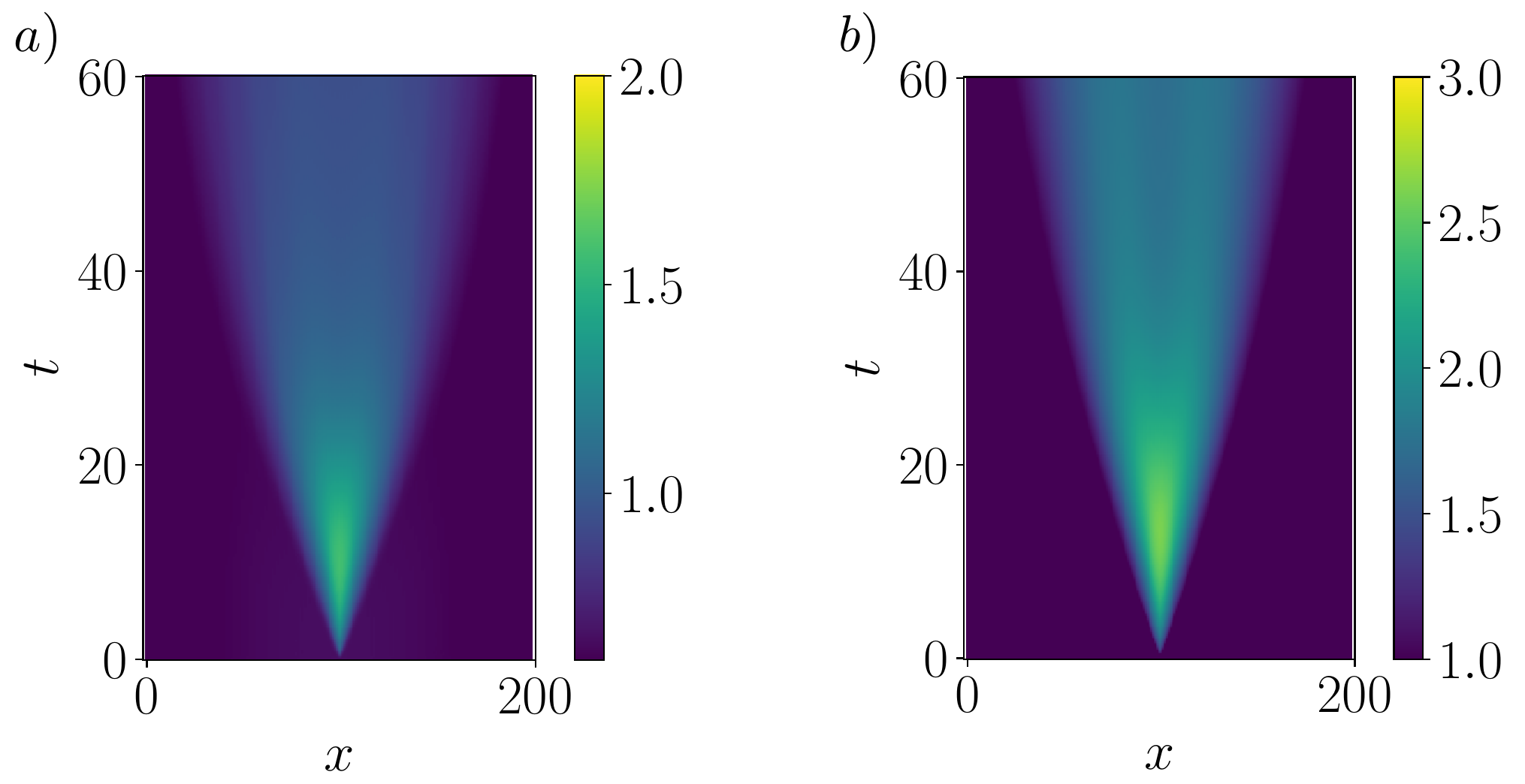}
    \caption{Entanglement entropy as a function of time and space for the time evolution of the wavefunction used to calculate the main-text dynamical structure factors: a) $\mathcal{S}_{\perp}(k,\omega)$ and b) $\mathcal{S}_{zz}(k,\omega)$.  We use the critical coupling $g_1=g_c$, $g_2=J=1$, $\Delta t=0.1$, $\chi=100$, $L=200$, and $T_{\text{max}}=60$.}
    \label{fig:ent_cone}
\end{figure}

\begin{figure*}[ht!]
    \centering
    \includegraphics[width=0.9\textwidth]{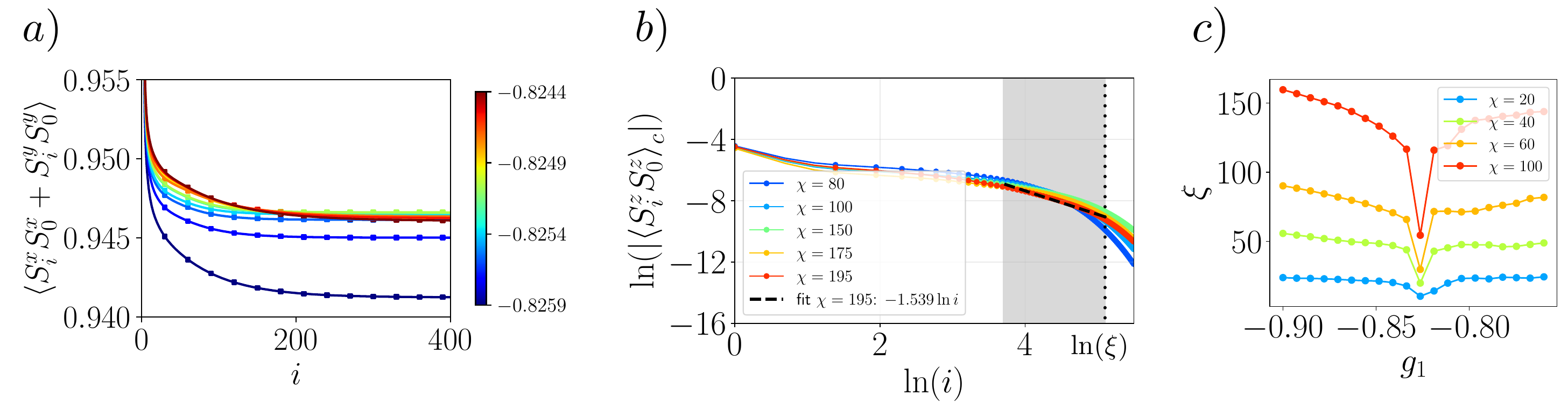}
    \caption{iDMRG results for different bond dimensions $\chi$ of the model in the main text with $g_2=1,J=1$.  We plot in a) the XY correlation function as a function of the number of sites for $g_1$ close to the critical point, b) Log-log plot of the equal-time connected correlation function for the $S^z_i$ component of the spin for a finite-size system with $L=200$. We plot the correlations for different values of $\chi$.  We find the relation given in Eq.~\eqref{corr_z} for the critical exponent. It is confirmed independently within the dynamical correlation function in the SM. The fit window shown in gray has as a lower bound a short-distance scale after which one expects long-distance physical effects. The upper bound is marked by a dotted line and is given by the maximum correlation length obtained from the iMPS for the bond dimension $\chi=195$. We plot in c) the correlation length obtained from iDMRG as a function of coupling $g_1$.}
    \label{fig:supp_plots}
\end{figure*}
For the dynamical structure factor we use a finite-chain time-evolution calculation.  We first obtain an iDMRG ground state, repeat and cut it to a finite chain of length $L$, and then perform additional finite DMRG sweeps.  After applying the local operator $S^\pm_0$ or $S^z_0$, the state is evolved with TEBD. Time-evolving block decimation (TEBD) is a tensor-network algorithm for simulating the real-time evolution of a matrix product state under a local nearest neighbor Hamiltonian. The time-evolution operator is decomposed into a sequence of two-site gates using a Trotter-Suzuki expansion. After each application of the local gates, the resulting state is compressed back to a finite bond dimension $\chi$, so the accuracy is controlled by the time step, the Trotter error, and the truncation error associated with finite $\chi$. The Fourier transform is performed with a Gaussian time window,

\begin{equation}
    C_{\alpha\beta}(r,t)\longrightarrow
    e^{-\sigma^2t^2}C_{\alpha\beta}(r,t),
\end{equation}
where we used $\sigma=0.2$. The finite time and finite system size impose the resolutions
\begin{equation}
    \delta \omega \sim \frac{2\pi}{t_{\max}},
    \qquad
    \delta k=\frac{2\pi}{L},
\end{equation}
in addition to the broadening from the Gaussian kernel.  Accessing the joint small-$k$, small-$\omega$ regime therefore requires increasing both $L$ and $t_{\max}$.  At the same time, the local perturbation produces an entanglement cone; once this cone reaches the boundary, the finite-chain response is affected by reflections.  We therefore explicitly check that the entanglement generated during time evolution remains away from the boundaries over the time window used for the Fourier transform, as shown in Fig.~\ref{fig:ent_cone}.

The leading cost of a two-site MPS update is dominated by the SVD and scales approximately as $\mathcal{O}(d^3\chi^3)$, where $d$ is the local Hilbert-space dimension and $\chi$ is the bond dimension. For a finite-chain TEBD calculation with $N_t=t_{\max}/\Delta t$ time steps and $\mathcal{O}(L)$ two-site gates per time step, the total cost therefore scales roughly as
\begin{equation}
    \mathcal{O}\!\left(L\,\frac{t_{\max}}{\Delta t}\,d^3\chi^3\right).
\end{equation}
In the convergence checks used here, typical parameters are $L=150$--$400$, $\chi=100$--$300$, and $\Delta t=0.1$ or $0.05$. Increasing $\chi$ from $150$ to $300$ increases the leading cubic cost by a factor of $2^3=8$, while halving $\Delta t$ doubles the number of time steps. This scaling explains the practical limitation in reaching simultaneously small momentum and small frequency: resolving small momenta requires large systems, $\Delta k\sim 2\pi/L$, while resolving small frequencies requires long real-time evolution, $\Delta\omega\sim 1/t_{\max}$, which is itself limited by entanglement growth and the finite accessible bond dimension.

\section{Extraction of critical exponents from static observables}
\label{sec:static_exponents} This section explains how the extraction of critical exponents from the static observables was performed. The ground-state quantities are computed by slowly increasing the bond dimension. After sufficiently many sweeps at a given low bond dimension, we save the final wavefunction and use it as the initial state for the next bond dimension.  We also use the largest-$\chi$ wavefunction at a fixed parameter value $g_1$ as the initial state for the next value $g_1^\ast=g_1+\delta g_1$, with $\delta g_1=0.007$.  This procedure ensures stability and allows us to monitor convergence with $\chi$. 

The longitudinal magnetization acts as an order parameter given in general by
\begin{equation}
    m_z=\dfrac{1}{L}\sum_i\langle S^z_i\rangle ,
\end{equation}
Close to the transition we expect to have for a second order phase transition the power-law behavior 
\begin{equation}
    m_z(g_1) = A\,(g_c-g_1)^\beta ,
    \qquad g_1<g_c.
    \label{eq:static_beta_fit}
\end{equation}
It is worth emphasizing that since iDMRG works in the thermodynamical limit explicitly and assumes translational invariance it means that $m_z$ reduces to 
$m_z= \expval{S^z_i}$ which is the same for all sites. We then perform a linear fit in log-log to determine the exponent $\beta$. Recalling that we tested for the stability as the bond dimension is increased we assume that we are extracting the physically correct $\chi\rightarrow \infty$ behavior.

\begin{figure}[t]
    \centering
    \includegraphics[width=\columnwidth]{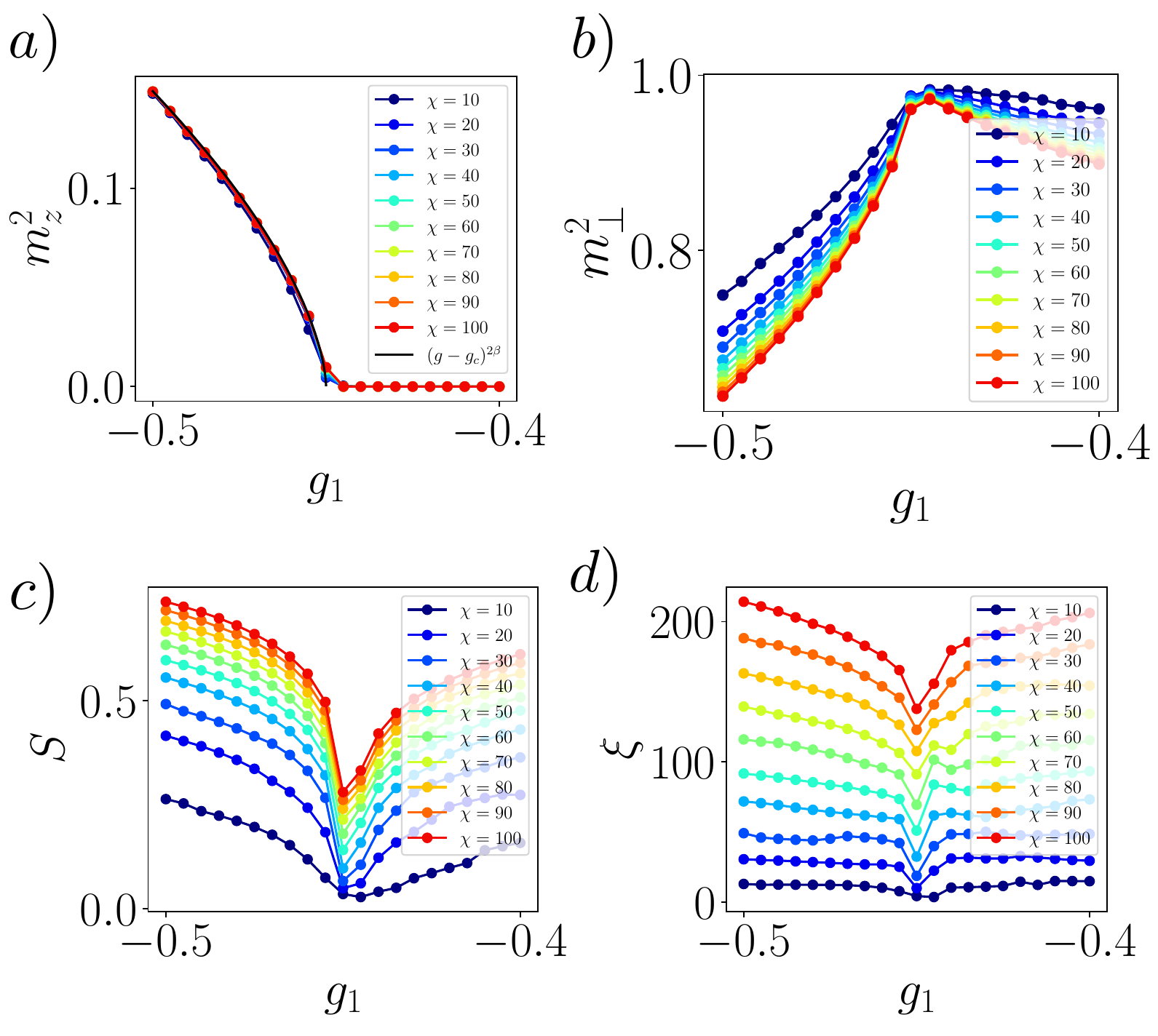}
    \caption{Numerical results using iDMRG for the main-text Hamiltonian with $g_2=0.5,J=1$, also showing a phase transition.  We plot a) the square of the order parameter $m_z$ for different bond dimensions $\chi$ and fit to a power law giving $\beta=0.31\pm0.01$, where the critical point is $g_c=-0.45\pm0.001$; b) perpendicular magnetization density for the same parameter values; c) entanglement entropy; and d) correlation length.}
    \label{fig:g2_0p5}
\end{figure}
For the critical exponent $\nu$ we can obtain it from either static or dynamical observables. We confirm that both methods converge to the same exponent up to the error indicated in the main text. We can extract $\nu$ from the power-law decay of the $S^z$ correlation functions. The correlation function of the $S^x,S^y$ (XY) operators instead shows the $U(1)$ long-range order at the critical point as seen We observe in Fig. ~\ref{fig:supp_plots} a) for $g_2=1$. As we approach the  critical point the XY correlation function changes from a power-law decay to an increasingly flat large-distance behavior. 
For Fig. ~\ref{fig:supp_plots} b) we introduce the connected longitudinal correlator
\begin{equation}
    C_z(x)=
    \langle S^z_x S^z_0\rangle
    -\langle S^z_x\rangle\langle S^z_0\rangle .
    \label{eq:connected_static_corr}
\end{equation}
At criticality we fit
\begin{equation}
    |C_z(x)| \sim x^{-2\Delta_\psi},
    \qquad
    \ln |C_z(x)| = \mathrm{const}-2\Delta_\psi\ln x ,
    \label{eq:static_corr_fit}
\end{equation}
over a window $x_{\min}<x<x_{\max}$. We recall here that $\Delta_\psi=(d+z+\eta-2)/2$ from which we obtain then a relationship between $z$ and $\eta$ mentioned in the main text. By calculating $z$ in the next sections we will obtain $\eta$. The lower cutoff removes short-distance lattice effects, this indicates the left boundary of the shaded area in Figure~\ref{fig:supp_plots} b). The upper cutoff avoids the finite-entanglement regime controlled by the correlation length $\xi$ that is itself a function of the bond dimension and is calculated from the iMPS form. We observe that this value is also the one for which the different bond dimension curves start to deviate from each other, as expected from finite bond-dimension effects. The correlation length in Fig. ~\ref{fig:supp_plots} c) increases with bond dimension on both sides of the transition, consistent with quasi-long-range order of the XY components away from the critical point. The special point $g_1=g_c$ displays not only the slowest decrease of the transverse magnetization with $\chi$ but also of the entropy and correlation length, which indicates the system fluctuates less strongly, consistent with $U(1)$ LRO. 

We also verify that the transition occurs away from the main value $g_2=1$.  Figure~\ref{fig:g2_0p5} shows iDMRG results for $g_2=0.5$.  The squared longitudinal order parameter is fitted to Eq.~\eqref{eq:static_beta_fit}, giving $g_c=-0.45\pm0.001$ and $\beta=0.31\pm0.01$ for the largest bond dimension used in that data set.  The transition in the longitudinal component occurs together with a maximum of the transverse magnetization density.  The accompanying increase of the correlation length and entropy indicates that the neighboring phases retain XY quasi-long-range order.  At the transition, the decrease of $m_\perp$ is the slowest, consistent with true XY LRO.

\section{Details on the numerical evidence for $U(1)$ SSB from static observables}
To study the $U(1)$ symmetry properties of the ground state at the critical point we use the squared of the transverse magnetization
\begin{align}
m_\perp^2=\tfrac{1}{L}\sum_i \expval{S_i^x}^2+\expval{S_i^y}^2,
\end{align}
It is worth noting that even though this parameter is squared it is composed of 1-point functions and thus detects transversal XY order. A non-zero value of this order parameter obtained with the iDMRG algorithm indicates spontaneous symmetry breaking 
\begin{align}
m^2_\perp\neq 0 \ \Rightarrow \ U(1) \; \text{SSB}.
\end{align}
We note that the translational invariance of the iMPS implies
\begin{align}
m_\perp^2= \expval{S_i^x}^2+\expval{S_i^y}^2,\  \forall \ i \quad \text{for} \ket{\psi} \text{an iMPS}.
\end{align}

The reason why a nonzero value of $m^2_\perp$ implies $U(1)$ SSB is simple. Consider $U_{\theta}$ the unitary that implements the $U(1)$ transformation by some angle $\theta$. Now if in the thermodynamic limit $m_\perp^2\neq 0$ as in the special point of the plot, then either $\expval{S_i^x}\neq 0$ or $\expval{S_i^y}\neq 0$, since is a sum of squares. We can now prove this implies SSB, we do it by contradiction. Assume the ground state  $\ket{GS}$ is not symmetry broken (is $U(1)$ symmetric), which implies 
\begin{align}
U_{\theta} \ket{GS}=\kappa_\theta \ket{GS},
\end{align}
 for some constant $\kappa_\theta$.  Now consider first the case $\expval{S_i^x}\neq 0$ then
 \begin{align}
 0\neq \expval{S_i^x}&= \expval{U_{\theta}^\dagger U_{\theta}S_i^xU_{\theta}^\dagger U_{\theta}}{GS}\\
 &= \abs{\kappa_\theta}^2 \expval{U_{\theta}S_i^xU_{\theta}^\dagger }{GS}.
\end{align}
But $U_{\theta}S_i^xU_{\theta}^\dagger$ rotates the spin operator, in particular for a 180 rotation $U_{\pi}S_i^xU_{\pi}^\dagger = -S_i^x$, therefore \begin{align}
\kappa^2_\pi \expval{U_{\pi}S_i^xU_{\pi}^\dagger }{GS} = -\kappa^2_\pi \expval{S_i^x }{GS},
\end{align}
implying $\expval{U_{\pi}S_i^xU_{\pi}^\dagger }{GS}=0$ or $\kappa_\pi =0 $ but $U_\pi$ is a unitary operator so it cannot eliminate states ($U_\pi\ket{GS}\neq 0$) therefore $\expval{U_{\pi}S_i^xU_{\pi}^\dagger }{GS}=0$ which contradicts that $\expval{S_i^x}\neq 0 $. The same argument applies for the $y$ component.
\begin{figure*}[t]
    \centering
    \includegraphics[width=0.8\textwidth]{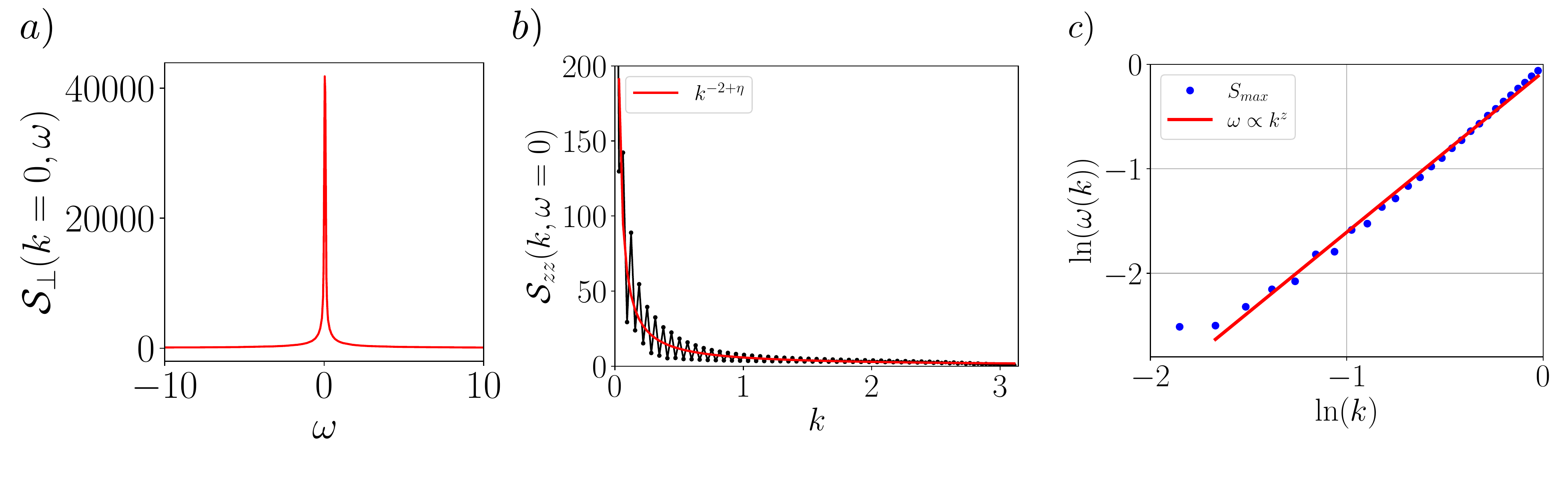}
    \caption{Dynamical structure factor cuts for the main-text plot: a) Bragg peak or condensed fraction shown in $\mathcal{S}_\perp(k=0,\omega)$; b) longitudinal dynamical structure factor as a function of momentum, fitted to a power law with $\eta=1.01\pm0.05$; c) power-law fit with slope $z=1.51\pm0.03$.  The points in c) are the centers of Gaussian fits to fixed-momentum cuts of the longitudinal spin structure factor.}
    \label{fig:Struct_cuts}
\end{figure*}

\section{Extraction of critical exponents from dynamical observables}
\label{sec:dynamical_exponents} We focus our analysis on the dynamical correlation function
\begin{equation}
    S_{\alpha\beta}(k,\omega)
    =
    \sum_r \int_{-\infty}^{\infty}\dd t\,
    e^{\ii(\omega t-kr)}
    \langle S^\alpha_r(t)S^\beta_0(0)\rangle .
    \label{eq:dsf_definition}
\end{equation}
The numerical time evolution described in Sec.~\ref{sec:tensor_network_methods} allows us to obtain both the transverse response $\mathcal{S}_\perp(k,\omega)$ and the longitudinal response $\mathcal{S}_{zz}(k,\omega)$. Fig. ~\ref{fig:Struct_cuts} shows cuts used to extract the low-energy spectral information for the bond dimension $\chi=100$. We will show  by doing a finite-size and entanglement scaling why this bond dimension captures the critical properties of larger bond-dimensions. Panel a shows the zero-momentum transverse peak expected for a system with a Bragg peak, displaying XY ordering. Panel b fits the zero-frequency momentum dependence of the longitudinal response to the scaling form $\mathcal{S}_{zz}(k,\omega=0)\propto k^{-(2-\eta)}$, giving $\eta=1.01\pm0.05$, this value gives an independent measure of $\eta$. It is important to remark that this value is consistent with the one obtained from the static correlation function which gave $\Delta_\psi$ together with the fitted value of $z$ which is shown in panel c. Fig. ~\ref{fig:Struct_cuts}c extracts the dispersion by taking fixed-$k$ cuts of $\mathcal{S}_{zz}(k,\omega)$. For each fixed momentum we fit the frequency functional form to a Gaussian
\begin{equation}
    I_k(\omega)
    =
    A_k\exp\!\left[-\frac{(\omega-\omega_k)^2}{2s_k^2}\right]
    +B_k ,
    \label{eq:gaussian_peak_fit}
\end{equation}
after this Gaussian fitting we plot the values of the frequencies against momentum and fit  
\begin{equation}
    \omega_k = a |k|^z ,
    \qquad
    \ln\omega_k=\ln a+z\ln|k| .
    \label{eq:z_fit}
\end{equation}

Repeating the fit over different intervals resulted in values concentrated near $1.5$, we used intervals of points inside $(\epsilon,1)$ and $\epsilon>0$. The values of the exponent were concentrated near $1.5$ with a range from $1.48$ for small $\epsilon$ to $1.56$ for larger $\epsilon$ and had error fits of approximately $0.02$ from which we extracted the value $z=1.51\pm 0.03$. 
\begin{figure}[b]
    \centering
    \includegraphics[width=\columnwidth]{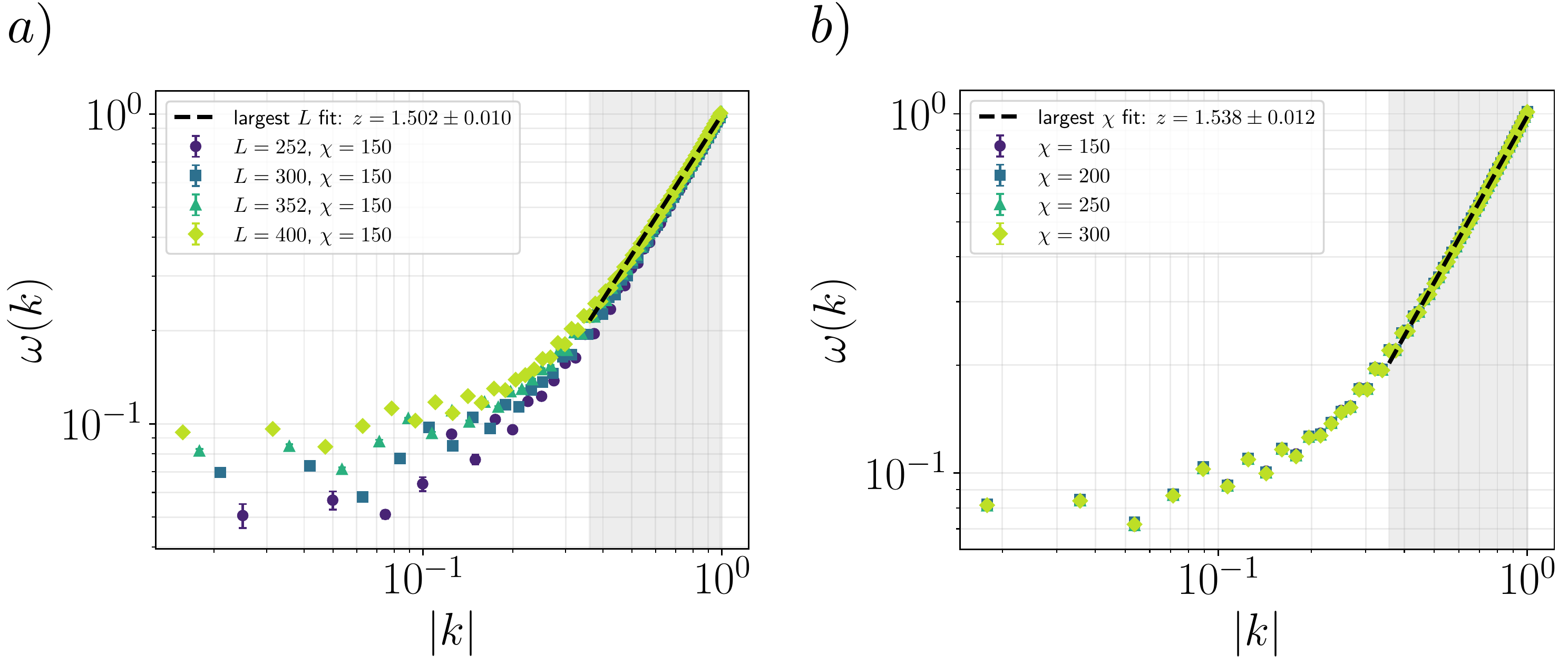}
    \caption{a) Centers of Gaussian fits to fixed-momentum cuts of the longitudinal spin structure factor for fixed bond dimension $\chi=150$, increasing length $L$ and maximum evolved time $t_{max}=75,90,105,120$. The fitting windows are displayed in gray. The upper limit is chosen so as to satisfy the long wavelength limit $k<1$ necessary to observe long-distance universal physics belonging to the critical point. The lower limit is chosen to be the smallest value of $k$ for which we have a scaling collapse as a function of $L$. In panel b) we perform a similar fit, in this case for a fixed length $L=352$ and maximum time $t_{max}=105$ for increasing bond dimensions.}
    \label{fig:dynamic_scaling_L_chi}
\end{figure}

\begin{figure}[t]
    \centering
    \includegraphics[width=\columnwidth]{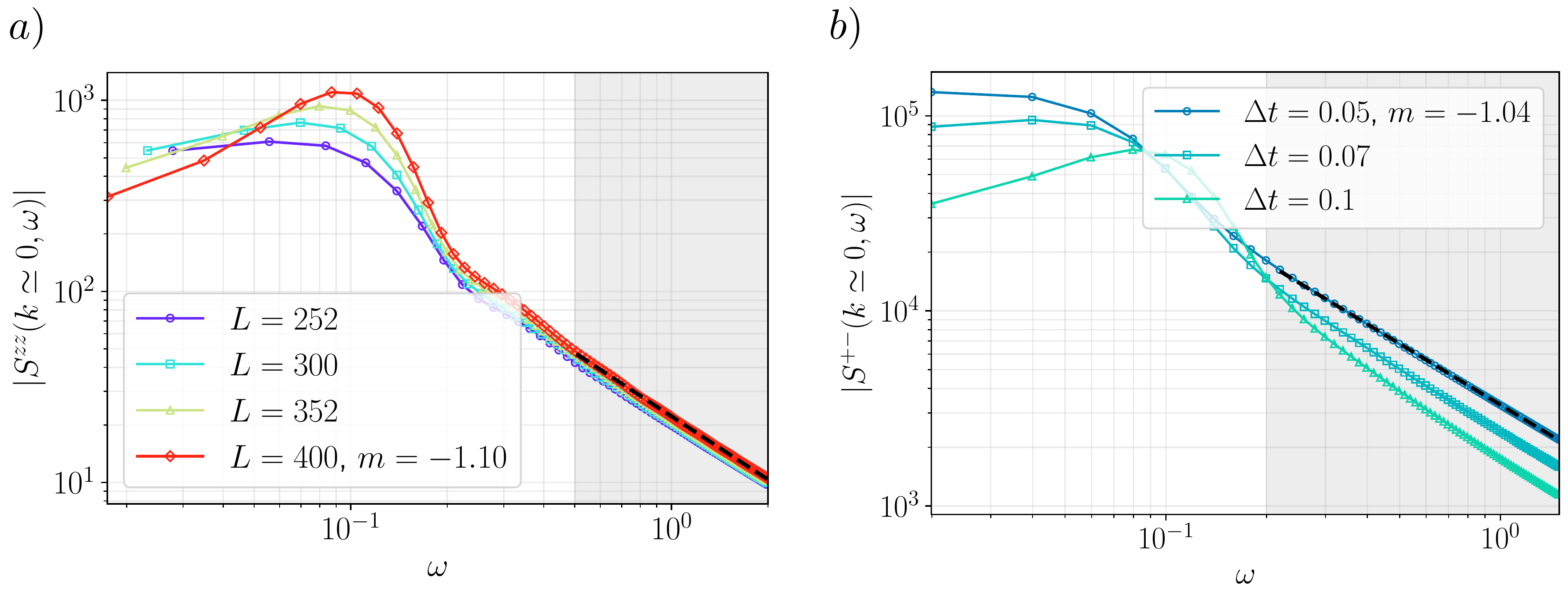}
    \caption{a) Log-log plot of the dynamical structure factor $S^{zz}(k\simeq0,\omega)$ cut at $k\simeq 0$ as a function of frequency for increasing system sizes and increasing maximum times at fixed $\chi=150$. We perform a fit for the largest system size displayed as a dashed line. In panel b we plot $S^{\perp}(k\simeq0,\omega)$ but for a fixed  $\chi=150$ and system size of $L=352$. The longitudinal channel is compared across system sizes, and the transverse channel is compared across Trotter steps. The gray windows indicates the frequency interval used for the log-log fit.  }
    \label{fig:k0_omega_scaling}
\end{figure}

\begin{figure*}[ht!]
    \centering
    \includegraphics[width=0.7\textwidth]{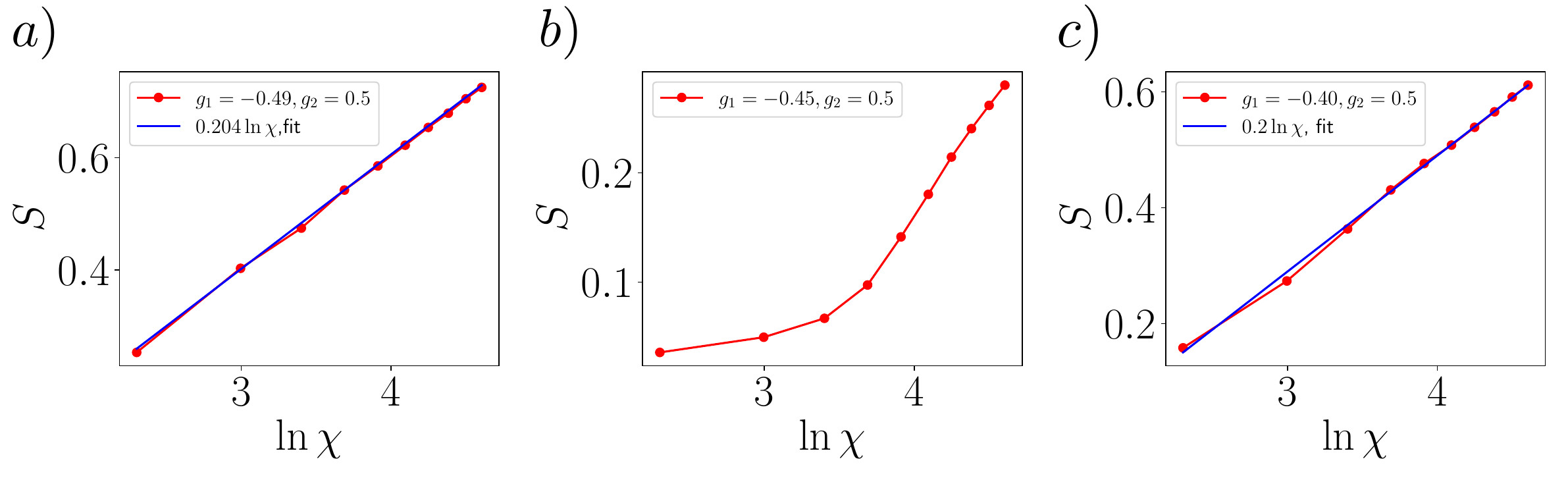}
    \caption{Entanglement entropy scaling as a function of bond dimension for couplings $g_2=0.5,J=1$.  We plot a) the scaling at the XY QLRO and Z LRO point $g_1=-0.49$, finding a logarithmic dependence with coefficient $\kappa=0.2$; b) non-logarithmic scaling at the critical coupling $g_1=g_c=-0.45$; and c) scaling at the XY QLRO and Z disordered point $g_1=-0.4$, with a similar logarithmic coefficient $\kappa=0.201$.}
    \label{fig:ent_entropy_scaling}
\end{figure*}
To further confirm that the dispersion is given by this dynamical critical exponent we further perform a finite-size scaling in $L$ and an increase in bond dimension $\chi$. Fig~\ref{fig:dynamic_scaling_L_chi} shows these checks, at fixed $\chi=150$, we increase the system size $L$ together with the maximum time $t_{max}$ such that the entanglement cone does not reach the boundaries of the system. The largest-size data set gives a fit to $z=1.502\pm0.010$ over the displayed fitting window. The lower limit is necessary since the finite $L,t_{max},\chi$ time evolution constrains us to a capped entanglement entropy and thus limits us as explained before to a finite momentum $k_{low}$ and frequency $\omega_{low}$. At fixed length, the largest bond dimension gives $z=1.538\pm0.012$. The spread between these estimates is the remaining finite-size and finite-entanglement systematic uncertainty resulting in an estimate of $z=1.50\pm 0.04$. It is worth emphasizing that the increase in bond dimension has very little effect on the $\omega,k$ points as seen in Fig.~\ref{fig:dynamic_scaling_L_chi} b. We take this as evidence that the lower bond dimensional wavefunction, used for other observables as in Fig.~\ref{fig:g2_0p5}, displays a very similar behavior as larger bond dimension wavefunctions.

We further analyze the convergence with increasing system size $L$ and decreasing Trotter step $\Delta t$ in Fig~\ref{fig:k0_omega_scaling} by plotting  $S^{zz}(k\simeq0,\omega)$ and $S^{\perp}(k\simeq0,\omega)$. This cut is shown as a high intensity vertical streak at $k=0$ in the main text dynamical structure factor figure. We argue with Fig~\ref{fig:k0_omega_scaling} that this is not an artifact of the method or the numerical approximations, but an expected behavior at criticality. Indeed, the log-log fit in both panels implies 
\begin{align}
S^{+-}(k\simeq0,\omega) \propto S^{zz}(k\simeq0,\omega) \propto\dfrac{1}{\omega^{\abs{m}}},\  \abs{m}= 1.1\pm 0.1 
\end{align} 
The power-law behavior is expected for critical systems, moreover the value of approximately one is consistent with other common spectral density distributions such as pink noise or fractal noise which is of the form $1/\omega^\alpha$ with $\alpha$ not an even integer. It is worth mentioning that the fitted value of $m$ is consistent for both larger sizes and smaller time indicating good convergence of the dynamical observables and critical properties of the ground state.  We remark also that as $\Delta t$ is decreased the power-law behavior as a function of $\omega$ persists for smaller values as seen by comparing $\Delta t=0.05$ and  $\Delta t=0.1$ in  Fig~\ref{fig:k0_omega_scaling}, but the exponent $m$ doesn't change for the values of frequencies where the functional form is a power-law. 

\section{Entanglement entropy scaling}\label{sec:entropy_scaling}

Finally, we discuss the finite-entanglement scaling. Within the matrix product state framework we have access to the half-chain von Neumann entropy defined by
\begin{equation}
    S_{\mathrm{vN}}=-\mathrm{Tr}\rho_A\ln\rho_A .
\end{equation}
We choose $A$ to be the half-chain subspace of the total Hilbert space In a one-dimensional conformal critical point, finite-entanglement scaling gives
\begin{equation}
    S_{\mathrm{vN}}\propto \frac{c}{6}\ln\xi_\chi ,
    \label{eq:entropy_xi_scaling}
\end{equation}
where $\xi_\chi$ is the correlation length for an iMPS that depends on the bond dimension used.Alternatively as a function of bond dimension,
\begin{equation}
    S_{\mathrm{vN}}\propto \kappa\ln\chi,
    \qquad
    \kappa=\frac{1}{\sqrt{12/c}+1},
    \label{eq:entropy_chi_scaling}
\end{equation}
up to finite-entanglement corrections~\cite{Hastings2007,Vidal2007,ScalingTagliacozzo,CumulantsiDMRG,stojevic_conformal_2015}.

The results are shown in Fig.~\ref{fig:ent_entropy_scaling}.  In panels a) and c), corresponding to the QLRO phases on the two sides of the transition for $g_2=0.5$, the entropy follows a logarithmic dependence on $\chi$ with $\kappa=0.2$ and $\kappa=0.201$, respectively.  This is consistent with a $c=1$ Gaussian-boson description of the neighboring XY quasi-long-range-ordered phases.  In contrast, the critical coupling $g_1=g_c=-0.45$ in panel b) does not follow the same simple logarithmic scaling.  Together with the dynamical evidence for $z\neq1$, this supports the interpretation of the transition as nonconformal.

\section{Derivation of the action}

In this section, we derive the action starting from the Hamiltonian of Eq.~\eqref{Ham}. We use the coherent state representation of the spin, defined by the equation $\expval{\textbf{S}}{\textbf{n}}=S\textbf{n}$, they form a basis $1=\tfrac{2S+1}{4\pi}\int \dd \Omega \ketbra{\textbf{n}}$ and non-linear constraint $\textbf{n}^2=1$ must be satisfied at all times and for all sites. We parametrize the components of the vector such that the constraint is always satisfied by $\textbf{n}=(\sqrt{1-\psi^2}\cos\theta,\sqrt{1-\psi^2}\sin\theta,\psi)$ with $\psi\in[-1,1]$ and $\theta\in (0,2\pi]$. Let us calculate the partition function of the quantum system by inserting identity operators at each discretized imaginary-time step. The usual procedure leads to the lattice action  
\begin{align}
    S=&\int_0^\beta \dd \tau \sum_{j=1}^N (-\ii S(1-\psi_j)) \partial_\tau \theta_j
    -JS^2 \textbf{n}_j(\tau)\cdot\textbf{n}_{j+1}(\tau)\notag\\&+g_1(S^2\psi_j^2+\frac{S}{2}(1-\psi^2_j))+g_2(S^2\psi_j^2+\frac{S}{2}(1-\psi^2_j))\notag\\
  & \qquad \times (S^2\psi_{j+1}^2+\frac{S}{2}(1-\psi^2_{j+1})),
\end{align}
where we defined $\textbf{n}_j(\tau)=\textbf{n}(x_j,\tau)$ the spin configuration at position $x_j=aj$ and $\Delta x_j=a$ is the lattice constant, we take $N$ lattice sites. The first term comes from computing overlaps of coherent states at different times and is known as a Berry phase. The Berry phase ensures that $S$ is a half-integer generically, in our case $S=1$, and we fix it in the following and drop constant terms. We now want a theory in the continuum $a\rightarrow 0$ and $N\rightarrow \infty$ while keeping $Na = L$. For this, we perform an expansion in the lattice constant $a$. We expect this to be a good approximation near a second-order phase transition where the correlation length is much larger than any lattice spacing. We express then $\textbf{n}_{j+1}(\tau)=\textbf{n}(x_j,\tau)+a\partial_x \textbf{n}(x_j,\tau)+\tfrac{a^2}{2}\partial_x^2 \textbf{n}(x_j)+O(a^3)$. We must enforce $\textbf{n}(x,\tau)^2=1$ at each space-time point. Furthermore, this implies $\partial_x\textbf{n}\cdot \textbf{n}=0$. We use this to write 
\begin{align}
    S&= \int_0^\beta \dd \tau a^{-1}\int^L_0 \dd x (-\ii(1-\psi)) \partial_\tau \theta -J \textbf{n}\cdot(\textbf{n}+\tfrac{a^2}{2}\partial_x^2 \textbf{n})
    \notag \\&+\tilde{g}
    _1\psi^2+\frac{g_2}{4}\psi^2 (\psi+a\partial_x \psi+\tfrac{a^2}{2}\partial_x^2 \psi)^2+O(a^3),
\end{align}
where we also expanded the last terms and defined $\tilde{g}_1=(g_1+g_2)/2$. Now we use the expression
\begin{align}
    \partial_x \textbf{n} = &\left(-\dfrac{\psi}{\sqrt{1-\psi^2}}\partial_x \psi \cos \theta -\sqrt{1-\psi^2}\sin \theta \partial_x \theta,\right. \notag\\  & \left.-\dfrac{\psi \partial_x \psi }{\sqrt{1-\psi^2}}\sin \theta +\sqrt{1-\psi^2}\cos \theta \partial_x \theta , \partial_x \psi\right).
\end{align}
Integrating by parts, we obtain to order $O(a^3)$
\begin{align}
    &S= \int_0^\beta \dd \tau a^{-1}\int^L_0 \dd x (-\ii(1-\psi)) \partial_\tau \theta -J + \dfrac{J}{2}a^2 (\partial_x\textbf{n})^2\notag \\
    &+\tilde{g}_1\psi^2+\frac{g_2}{4}\psi^2 (\psi^2+2a \psi \partial_x \psi+a^2(\partial_x \psi)^2+a^2 \psi \partial_x^2 \psi)
\end{align}
After simplifications and dropping total derivatives and constant energy contributions, we obtain 
\begin{align}
     S= \int_0^\beta \dd \tau& a^{-1}\int^L_0 \dd x \; \ii \psi \partial_\tau \theta + \dfrac{J}{2}a^2\left( (\partial_x\psi )^2+\dfrac{\psi^2 (\partial_x \psi)^2}{1-\psi^2}\right.\notag \\
     &+(1-\psi^2)(\partial_x \theta)^2 +\tilde{g}_1\psi^2\notag \\
     &+\frac{g_2}{4}\psi^2 (\psi^2+a^2(\partial_x \psi)^2+a^2 \psi \partial_x^2 \psi)
\end{align}
Final expression for the Lagrangian with $S=1=a$, dropping linear terms in the fields and everything is to order $\mathcal O(a^2)$:
\begin{align}
    \mathcal L = \ii &\psi\partial_{\tau}\theta + \frac J 2 \Big[\frac{1}{1-\psi^2} (\partial_x\psi)^2+\big(1-\psi^2\big)(\partial_x\theta)^2\Big]\notag
\\
&+\tilde{g}_1\,\psi^2+\frac{g_2}{4}\,\psi^2\Big(\psi^2+(\partial_x\psi)^2+ \psi\partial_x^2\psi\Big)\;.
\end{align}
last term $\psi^3\partial_x^2\psi=-3\psi^2(\partial_x\psi)^2$, therefore, this 
to
\begin{align}
    \mathcal L = \ii \psi\partial_{\tau}\theta + \frac J 2 \Big[\frac{1}{1-\psi^2} (\partial_x\psi)^2+\big(1-\psi^2\big)(\partial_x\theta)^2\Big]\notag\\+\tilde{g}_1\,\psi^2+\frac{g_2}{4}\,\psi^2\Big(\psi^2-2(\partial_x\psi)^2\Big)\;.
\end{align}
Integrating out the phase leads to
\begin{align}
    \mathcal L = \frac 1 {2J} \frac{\psi\partial_{\tau}^2\nabla^{-2}\psi}{1-\psi^2}  + \frac J 2 \frac{(\partial_x\psi)^2}{1-\psi^2} +\tilde{g}_1\,\psi^2\notag\\
    +\frac{g_2}{4}\,\psi^2\Big(\psi^2-2(\partial_x\psi)^2\Big)\;.
\end{align}
Following the argument of Landau-Ginzburg, we drop higher order $\psi$-terms since it is the order parameter and it becomes small at the second order phase transition $\psi^2\approx0$
\begin{align}
    \mathcal L = \Big[\tfrac 1 {2J} \psi\partial_{\tau}^2\nabla^{-2}\psi  + \tfrac J 2 (\partial_x\psi)^2\Big](1+\psi^2) +\tilde{g}_1\,\psi^2\notag \\
    +\frac{g_2}{4}\,\psi^2\Big(\psi^2-2(\partial_x\psi)^2\Big)+\mathcal O(\psi^6)\;.
\end{align}
Rewriting the result, we propose the following initial Lagrangian
\begin{align}
    \mathcal L = \tfrac 1 {2} \psi\partial_{\tau}^2\nabla^{-2}\psi  + \tfrac 1 2 (\partial_x\psi)^2
    +m\,\psi^2
    + \lambda_0 \psi^4 \notag
    \\
    + \lambda_1 \psi^3\partial_{\tau}^2\nabla^{-2}\psi  + \lambda_2 \psi^2 (\partial_x\psi)^2 \;.
\end{align}
Compared to the Lagrangian $\mathcal L_1$ in the main text, there are two additional terms
\begin{align}
    \mathcal L_1 + (\tfrac J 2-\tfrac{g_2}{2}) \psi^2 (\partial_x\psi)^2-\tfrac J2 \psi^2(\partial_x\theta)^2\;.
\end{align}
Even though the Lagrangian is still quadratic in $\theta$, there are now bare four-point vertices containing $\theta$, such that the second Ward identity no longer holds. This leads to more diagrams that have to be taken into account at a given order and might change the dynamics and critical exponents substantially.\\

%%%%%%
\section{Epsilon expansion}
The bare massless Euclidean action (in imaginary time $\tau$) is given by
\begin{equation}
	\label{eq:S0_D}
	S_0[\theta,\psi]
	= \int \dd^d x\, \dd \tau \;
	\Big[
	\, \ii\,\, \psi\, \partial_\tau \theta
	+ \tfrac{K}{2}\,(\nabla\theta)^2
	+ \tfrac{1}{2}\, (\nabla\psi)^2
	+ \tfrac{g}{4!}\,\psi^4
	\Big]\;.
\end{equation}
With the Euclidean Fourier convention
\begin{align}
	f(\omega,p) = \int \dd^d x \dd \tau \; f(\tau,x)\, \e{-\ii\omega \tau-\ii p \cdot x}\;,
\end{align}
the bare inverse propagator is given by
\begin{align}
	S^{(2)}(\omega,q)=
	\begin{pmatrix}
		q^2 & \omega\\
		-\omega & K\,q^2
	\end{pmatrix}\;,
\end{align}
with a field ordering $(\psi,\theta)$. The bare Green's function for the $\psi$-fields $G_{0,\psi\psi}$, we will denote as $G_{0}(\omega,q)
=\frac{K\,q^2}{\omega^2 + K \,q^4\,}\,$. The poles are $\omega_{1,2}=\pm i\,\sqrt{K}\;q^2$.

\subsection{Multiplicative renormalization}
We introduce renormalized fields and couplings, denoted by a subscript $R$. We closely follow the derivation in Zinn-Justin’s book~\cite{ZinnJustin_book} and adopt its notation where convenient. The scale $\mu$ is the renormalization scale, i.e. the symmetric momentum scale at which we evaluate the momenta of the 1PI correlation functions.
\begin{equation}
	\theta = Z_\theta^{1/2}\, \theta_R,\;
	\psi  = Z_\psi^{1/2}\, \psi_R\,, \;
	K   = \frac{Z_{K}}{Z_{\theta}} \, K_R   \;, \;
	g   = \frac{\mu^{\varepsilon}\, Z_{g}}{Z_{\psi}^2}\, g_R\;.
\end{equation}
In MS with dimensional regularization, only poles in \(\varepsilon\) are subtracted in the \(Z\)–factors.
\begin{align}
	\label{eq:SR_D}
	&S_R[\theta_R,\psi_R]
	= \int \dd^d x\, \dd \tau \;
	\Big[
	\, \ii\,\big(Z_\psi^{1/2} Z_\theta^{1/2}\big)\, \psi_R\, \partial_\tau \theta_R \notag
	\\&+ \frac{1}{2}\,Z_K \, K_R\, (\nabla\theta_R)^2
	+ \frac{1}{2}\,Z_\psi\, (\nabla\psi_R)^2
	+ \frac{1}{4!}\,\mu^{\varepsilon}\,Z_g \, g_R\, \psi_R^4
	\Big].
\end{align}

% --- Canonical (Berry) normalization / Ward constraint in this convention ---
\subsection{Ward identities}
The nonrenormalization of the Berry term leads to a Ward identity for the mixed two-point vertex at the normalization point: $\partial_\omega\Gamma^{\psi\theta}_{R}\big|_{\text{NP}}=1$. This imposes the constraint among Z-factors:
\begin{equation}
	Z_\psi^{1/2}\,Z_\theta^{1/2} = 1\;,
	\quad\Longleftrightarrow\quad
	Z_\psi =Z_\theta^{-1} \equiv Z\;.
\end{equation}
Moreover, the $\theta$ sector receives no independent vertex corrections because $\theta$ does not enter the interaction term. Equivalently, no 1PI diagram with external $\theta$ legs can be constructed from the quartic $\psi$ vertex. Although the connected propagator $G^{\theta\theta}$ is modified through its coupling to the interacting $\psi$ sector, the amputated vertex $\Gamma^{\theta\theta}=S^{\theta\theta}$ remains equal to its bare form. Consequently, the coefficient of $(\nabla\theta)^2$ is not independently renormalized, giving $Z_K=1$ and
\begin{align}
    Z\, K_R = K\;.
\end{align}
Note that both Ward identities can be derived from $\langle J_{\theta}\epsilon\rangle=\langle\delta S \rangle$ with the source field of the phase $J_{\theta}=\Gamma^{\theta}$, by noting that an infinitesimal space-time dependent shift of the phase
\begin{align}
	\theta(x,\tau)\mapsto\theta(x,\tau)+\epsilon(x,\tau)\;,
\end{align}
varies the action to first order in the shift
\begin{align}
	\delta S = \ii \psi\partial_{\tau}\epsilon +K\nabla\theta\nabla\epsilon+\mathcal{O}(\epsilon^2)\;.
\end{align}

\subsection{Ansatz}
The effective action, therefore, reads
\begin{align}
	\label{eq:SR_D2}
	&S_R[\theta_R,\psi_R]
	= \int \dd^d x\, \dd \tau \;
	\Big[
	\, \ii\, \psi_R\, \partial_\tau \theta_R
	\notag \\
    &+ \frac{1}{2} \, K_R\, (\nabla\theta_R)^2
	+ \frac{1}{2}\,Z\, (\nabla\psi_R)^2
	+ \frac{1}{4!}\,\mu^{\varepsilon}\,Z_g \, g_R\, \psi_R^4
	\Big].
\end{align}
Hence, the only independent renormalization constants are those of the massless $\phi^4$-theory; the running of $K_R$, $\mu\partial_\mu\ln K_R=-\eta$, is fixed by $Z$. From the Berry phase term can be seen that $\psi$ and $\theta$ are conjugate fields; moreover, their renormalization are locked together due to the first Ward identity.

\subsection{Massive theory}
In order to introduce a mass to the theory, we add a source term for the monomial $\psi^2_R$ to the action. We write
\begin{align}
	S_R[\theta_R,\psi_R,m_R] = S_R[\theta_R,\psi_R]+\frac{1}{2} m_R^2 \int_{x,\tau} Z_2 \psi_R^2\;,
\end{align}
where with respect to the initial action $m^2 = m_R^2\,Z_2\,Z^{-1}$.

\section{One-loop correction}
To one-loop order, we need to calculate only a single diagram that contributes at finite external momentum. This is the diagram $B_d(\omega,q)$ that we evaluate and expand in $\varepsilon$ about the upper critical dimension $d_c=2$ in Section~\ref{sec_diagramB}.

\subsection{No correction to the field renormalization}
First, we see that at one-loop order, all Z factors of terms involving derivatives remain one. In particular, the field renormalization factor satisfies
\begin{align}\label{1loop_Z}
	Z = 1 + \mathcal O(g^2_R)\;. 
\end{align}

\subsection{Four-point function}
We calculate here the three channels of the diagram $B_d(q)$, see the appendix, and arrive at one loop at $\mathcal O(g_R^3)$
\begin{align}
	\tilde \Gamma^{(4)}_R(p_i) &= g_R Z_g \mu^{\varepsilon}-\frac{1}{2} Z_g^2 g_R^2 \mu^{2\varepsilon}\Big[B_d(p_1+p_2)+2 \text{ channels}\Big]\;,
\end{align}
inserting the result at $\mu=1$, we obtain in $d=2-\varepsilon$ dimensions in the $\overline{MS}$ scheme
\begin{align}
	&\tilde \Gamma^{(4)}_R(p_i) = g_R Z_g -\frac{3}{2} Z_g^2 g_R^2 b_1 \\
    &- \frac{1}{2} Z_g^2 g_R^2 \Big[B_r(p_1+p_2)+2 \text{ channels}\Big] + \mathcal O(\varepsilon, g_R^3)\;,
\end{align}
with $B^{(\overline{MS})}_{2-\varepsilon}(q) = b_1+B_r(q)+\mathcal O(\varepsilon)$, $b_1\coloneqq\frac{\sqrt{K_R}}{4} N_2 \frac{1}{\varepsilon}$ being the divergent part in $\varepsilon$, but constant in momentum, and $B_r(q)\coloneqq-\frac{\sqrt{K_R}}{4} N_2\, [\frac{1}{2}\ln q^2-\ln 2]$ contains the IR divergence.
Therefore,
\begin{align}\label{1loop_Zg}
	Z_g \,=\,1\,+\,\frac{3}{2}\frac{\tilde g}{\varepsilon} + \mathcal O({\tilde g^2})\;,
\end{align}
where we introduce an effective coupling, absorbing the loop factors
\begin{align}
	\tilde g \, \coloneqq\, \frac{\sqrt{K_R}}{4} N_2 \, g_R\;.
\end{align}

\subsection{$\psi^2_R$-insertion}
We calculate the mixed three-point function
\begin{align}
	\tilde \Gamma^{(2,1)}_R(p_1,p_2) = Z_2 - \frac{1}{2} g_R Z_g Z_2 B_d(p_1+p_2) + \mathcal O(g_R^2)\;,
\end{align}
where the first index is a second-order derivative with respect to the $\psi_R$-field and the second index is the first derivative in the composite field $\psi_R^2$. We find at $p_2=0$ to one-loop order
\begin{align}
	\tilde \Gamma^{(2,1)}_R(p,0) = Z_2 - \frac{1}{2} g_R Z_g Z_2 (b_1+B_r(p)) + \mathcal O(\varepsilon,g_R^2)\;,
\end{align}
and find that
\begin{align}\label{1loop_Z2}
	Z_2\,=\,1\,+\,\frac{1}{2} \frac{\tilde g}{\varepsilon} \,+\,\mathcal O(\tilde g^2)\;.
\end{align}

\section{Two-loop order}
To two-loop order, we need to calculate three different diagrams, conveniently called $A$, $B$, and $C$. To enhance the readability of the derivation, the evaluation of the frequency and momentum integrals, as well as the expansion about the upper critical dimension, has been outsourced to Section~\ref{sec_diagramA},~\ref{sec_diagramB}, and ~\ref{sec_diagramC}.
\subsection{Two-point function}
We start with the two-loop correction to the $1$-PI self-energy of the renormalized two-point function in Fourier representation
\begin{align}
	\tilde\Gamma^{(2)}_R(q) = Zq^2-\frac{1}{6}g_R^2\mu^{2\varepsilon}A_d(q)+\mathcal O(g_R^3)\;.
\end{align}
that gets a correction from the setting sun diagram $A_d$. With the result~\eqref{diagramA_MS}. 
We can read off that the field renormalization becomes at $\mu=1$
\begin{align}\label{2loop_Z}
	Z\,=\,1\,-\,\frac{1}{54} \frac{1}{\varepsilon}\tilde g^2 \,+\,\mathcal O (\tilde g^3)\;.
\end{align}

\subsection{Four-point function}
To the next order, we can obtain the correction to $Z_g$ from the four-point function at two-loop order, comprised of
\begin{align}
	\tilde{\Gamma}^{(4)}_R(\omega_i,p_i)
	&= g_R\,Z_g\,\mu^{\varepsilon}\nonumber\\
	&\quad-\tfrac{1}{2}\,g_R^{2} Z_g^{2}\,\mu^{2\varepsilon}\,
	\bigg[\,\begin{aligned}
	&B_d(\omega_1+\omega_2,p_{1}+p_{2})\\
	&+ 2\ \text{terms}
	\end{aligned}\,\bigg]\nonumber\\
	&\quad+ \tfrac{1}{4}\,g_R^{3}\,\mu^{3\varepsilon}\,
	\bigg[\,\begin{aligned}
	&B_d^{2}(\omega_1+\omega_2,p_{1}+p_{2})\\
	&+ 2\ \text{terms}
	\end{aligned}\,\bigg]\nonumber\\
	&\quad+ \tfrac{1}{2}\,g_R^{3}\,\mu^{3\varepsilon}\,
	\bigg[\,\begin{aligned}
	&C_d(\omega_1,\omega_2,p_{1},p_{2})\\
	&+ 5\ \text{terms}
	\end{aligned}\,\bigg]\nonumber\\
	&\quad+ \mathcal{O}(g_R^{4})\,.
\end{align}
After setting $\mu=1$ and splitting $B_d(p)=b_{1}+B_{r}(p)$ as given in~\eqref{diagramB_MS_split}, where $B_{r}(p)$ is a constant in $\varepsilon$, the expansion becomes
\begin{align}
	\tilde{\Gamma}^{(4)}_R(p_i)
	&= g_{R}\,Z_g
	-\frac{3}{2}\,b_{1}\,g_{R}^{2}\,Z_g^{2}
	\nonumber\\
	&\quad-\frac{1}{2}\,Z_g^{2}\,g_{R}^{2}\,
	\big[\,B_{r}(p_{1}+p_{2})+2~\text{terms}\,\big]
	\nonumber\\
	&\quad+\frac{1}{4}\,g_{R}^{3}\,
	\big[\,B_{r}^{2}(p_{1}+p_{2})+2~\text{terms}\,\big]
	\nonumber\\
	&\quad+\frac{1}{2}\,b_{1}\,g_{R}^{3}\,
	\big[\,B_{r}(p_{1}+p_{2})+2~\text{terms}\,\big]
	\nonumber\\
	&\quad+\frac{3}{4}\,b_{1}^{2}\,g_{R}^{3}
	\nonumber\\
	&\quad+\frac{1}{2}\,g_{R}^{3}\,
	\big[\,C_d(p_{1},p_{2})+5~\text{terms}\,\big] \,.
\end{align}
We use the result of $Z_g$ to one-loop order~\eqref{1loop_Zg} to simplify perturbative expansion and extract the $Z_g$ to next order
\begin{align}
	\tilde{\Gamma}^{(4)}_R(p_i)
	&= g_{R}\,Z_g
	-\frac{3}{2}\,b_{1}\,g_{R}^{2}
	-\frac{15}{4}\,b_{1}^{2}\,g_{R}^{3}
	\nonumber\\
	&\quad-\frac{1}{2}\,g_{R}^{2}\,
	\big[\,B_{r}(p_{1}+p_{2})+2~\text{terms}\,\big]
	\nonumber\\
	&\quad+\frac{1}{4}\,g_{R}^{3}\,
	\big[\,B_{r}^{2}(p_{1}+p_{2})+2~\text{terms}\,\big]
	\nonumber\\
	&\quad+\frac{1}{2}\,g_{R}^{3}\,
	\bigg[\,\begin{aligned}
	&C(p_{1},p_{2})-b_{1}B_{r}(p_{1}+p_{2})\\
	&+5~\text{terms}
	\end{aligned}\,\bigg] \,.
\end{align}
Now, we make use of the fact that the IR-divergence in diagram $C_d$ comes from the subdiagram $B_d$, which cancels exactly. The divergent part of the combination of the diagrams $\big[C_d(p_{1},p_{2})-b_{1}\,B_{r}(p_{1}+p_{2})\big]_{2-\varepsilon}$ is therefore momentum–independent and given by
\begin{align}
	D_{\text{div}}
	&=
	C_{2-\varepsilon}^{(\overline{MS})}(q) - b_1\,B_r(q) \nonumber\\
	&= \frac{K_R\,N_2^2}{16}\Bigg\{
	\frac{1}{8}\Big[
	\frac 4 {\varepsilon^{2}}
	+(\tfrac{4}{3}+12\ln 2-2\ln 3
	- 4\ln q^2)\frac{1}{\varepsilon}
	\Big]\nonumber\\
	&+\frac{1}{\varepsilon}\left[\frac{1}{2}\ln q^2-\ln 2\right]
	+\mathcal O(\varepsilon^0)
	\Bigg\}\nonumber\\
	&= \frac{K_R\,N_2^2}{16}\frac{1}{8}
	\left[
	\frac 4 {\varepsilon^{2}}
	+(\tfrac{4}{3}+4\ln 2-2\ln 3 )
	\frac{1}{\varepsilon}
	\right]+\mathcal O(\varepsilon^0)\;.
\end{align}
Hence, the divergent part of $\Gamma^{(4)}$ reduces to
\begin{align}
	\big[\tilde{\Gamma}^{(4)}_{R}(p_i)\big]_{\text{div}}
	&= g_{R}\,Z_g
	-\frac{3}{2}\,b_{1}\,g_{R}^{2}
	-\frac{15}{4}\,b_{1}^{2}\,g_{R}^{3}
	+3\,g_{R}^{3}\,D_{\text{div}}\,.
\end{align}
We can read off
\begin{align}\label{2loop_Zg}
	Z_g &= 1 +\frac 3 2 b_1 \,g_R
	+ \frac{15}{4} b_1^2 \, g_R^2
	-3g_R^2\,D_{\text{div}}\\
	&= 1 + \frac 32 \frac{\tilde g}{\varepsilon}
	+ \tilde g^2\Bigg[
	\frac{9}{4}\frac{1}{\varepsilon^2}
	-\frac{1}{\varepsilon}\Big(\frac 12 +\frac 32\ln 2
	-\frac{3}{4}\ln 3\Big)\Bigg]
	\nonumber\\
	&\quad+\mathcal O(\tilde g^3, \varepsilon^0)\;.
\end{align}

\subsection{$\psi^2_R$-insertion}
We can calculate the correction to the mass from the three-point function at two-loop order
\begin{align}
	\tilde{\Gamma}^{(2,1)}_R(\omega_1,\omega_2,p_1,p_2)
	&= Z_2\nonumber\\
	&\quad-\tfrac{1}{2}\,g_R\,Z_g Z_2\,
	B_d(\omega_1+\omega_2,p_1+p_2)\nonumber\\
	&\quad+\tfrac{1}{4}\,g_R^{2}\,
	B_d^{2}(\omega_1+\omega_2,p_1+p_2)\nonumber\\
	&\quad+\tfrac{1}{2}\,g_R^{2}\,
	C_d(\omega_1,\omega_2,p_1,p_2)
	+\mathcal{O}(g_R^{3})\,.
\end{align}
Analogous to the previous section, we split $B_d$ and $C_d$, given in~\eqref{diagramB_MS_split} and ~\eqref{diagramC_MS} respectively, again into momentum-independent, divergent terms and momentum-dependent ones, insert the result for $Z_g$ and then $Z_2$ at one-loop order, and find the divergent part of the three-point function to be given by
\begin{align}
	\big[\tilde{\Gamma}_{R}^{(2,1)}(p,0)\big]_{\mathrm{div.}}
	&= Z_{2}
	- \tfrac{1}{2} b_{1} g_{R}\nonumber\\
	&\quad- \tfrac{3}{4} b_{1}^{2} g_{R}^{2}
	+ \tfrac{1}{2} g_{R}^{2}D_{\text{div}}
	+ O(g_{R}^{3}) \;.
\end{align}
Requiring that this amputated Green's function vanish at two-loop order leaves us with
\begin{align}\label{2loop_Z2}
	Z_2 &= 1+ \frac 12 \frac{\tilde g}{\varepsilon}
	+ \tilde g^2 \Bigg[
	\frac 12 \frac 1{\varepsilon^2}
	-\frac 1 {16} \big(\frac{4}{3}+ 4\ln2 -2\ln 3\big)
	\frac{1}{\varepsilon}\Bigg]
	\nonumber\\
	&\quad+\mathcal O(\tilde g^3 , \varepsilon^0)\;.
\end{align}

\section{Critical exponents}
\label{sec_crit_exp}
The dimension of the renormalized field is given by $[\psi_R]_{\mu}=\Delta_{\psi}=\tfrac12 (D-2+\eta)$ with $D=d+z$. The anomalous exponent is defined as the scaling of $Z$ at the fixed point
\begin{align}
	\eta\equiv\eta(\tilde g^*) \coloneqq \mu\partial_{\mu} \ln Z\big|_*\;,
\end{align}
where $\mu\partial_{\mu}\ln Z=\tilde\beta(\tilde g)\tfrac{\dd}{\dd\tilde g}\ln Z(\tilde g)$.
Since $K=Z\,K_R$, the renormalized coefficient runs, $\mu\partial_\mu\ln K_R=-\eta$,
and only the combination $\tilde g=\frac{\sqrt{K_R}}{8\pi}\,g_R$ reaches a fixed point.
The bare coupling then reads
\begin{align}
	g\sqrt{K}=\mu^{\varepsilon}\,\frac{Z_g}{Z^2}\,g_R\sqrt{Z\,K_R}
	=8\pi\,\mu^{\varepsilon}\,G(\tilde g)\;,
\end{align}
with $G(\tilde g)=\tilde g\,Z_g\,Z^{-3/2}$ and which gives at fixed bare couplings
\begin{align}
	\tilde\beta(\tilde g)\equiv\mu\partial_\mu\tilde g
	=-\varepsilon\left(\frac{\dd\ln G(\tilde g)}{\dd\tilde g}\right)^{-1}\;.
\end{align}
Further, we define $\eta_2(\tilde g)\coloneqq\tilde\beta(\tilde g)\tfrac{\dd}{\dd\tilde g}\ln\tfrac{Z_2(\tilde g)}{Z(\tilde g)}$,
which gives the critical exponent $\nu=(\eta_2(\tilde g^*)+2)^{-1}$.

Integrating out the phase field $\theta$, we find the renormalized action
\begin{align}
	\tilde S_R[\psi_R]
	= \int \dd^d x\, \dd \tau \;
	\Big[\,
	&\frac{1}{2\,K_R} \psi_R\, \partial_\tau^2 \nabla^{-2} \psi_R
	\nonumber\\
	&+ \frac{1}{2}\,Z\, (\nabla\psi_R)^2
	+ \frac{1}{4!}\,\mu^{\varepsilon}\,Z_g \, g_R\, \psi_R^4
	\,\Big].
\end{align}
The first term receives no counterterm, but its coefficient runs,
$\mu\partial_\mu\ln K_R=-\eta$. Since frequencies enter only through
$\omega/\sqrt{K_R}$, the characteristic frequency at momentum scale $\mu$ is
$\omega\sim\sqrt{K_R(\mu)}\,\mu^{2}\propto\mu^{2-\eta/2}$ at the fixed point, so that
the dynamical critical exponent is fixed by $\eta$,
\begin{align}\label{eq:scaling_z}
	z\,=\,2-\,\frac\eta 2\;.
\end{align}
Equivalently, the field dimension satisfies $2\Delta_\psi=d-z+2$, consistent with
$\Delta_\psi=\tfrac12(d+z-2+\eta)$.

\subsection{To first order}
We find from the first-order result that the $\beta$-function is given by
\begin{align}
    \tilde \beta (\tilde g)
	&= -\varepsilon\tilde g + \frac{3}{2} \tilde g^2
	+ \mathcal O(\tilde g^3)\;,
\end{align}
equivalent to the Ising one and find the non-trivial fixed point at $\tilde g^*= \frac{2}{3}\varepsilon$.

\subsection{To second order}
We get
\begin{align}
	\tilde \beta (\tilde g) &= -\varepsilon \tilde g
	+ \frac 32 \tilde g^2+\gamma_1\tilde g^3
	+ \mathcal O(\tilde g^4)\; \nonumber\\
	&\qquad \text{with}\quad
	\gamma_1\coloneqq-\frac{17}{18} -3\ln 2+\frac 32 \ln 3\;,\\
	\eta(\tilde g) &= \frac{1}{27} \tilde g^2\,+\,\mathcal O(\tilde g^3)\;,\\
	\eta_2(\tilde g) &= -\frac{\tilde g} 2 + \gamma_2 \tilde g^2
	+ \mathcal O (\tilde g^3)\;,
	\nonumber\\
	&\qquad \text{with}\quad
	\gamma_2
	\coloneqq
	\frac{7}{54}
	+\frac12\ln 2
	-\frac14\ln 3\;,
\end{align}
which is different from the correction in the Ising model. The positive, non-zero fixed point solution is given to second order in epsilon by $\tilde g^*=\frac 23 \varepsilon-\frac 8 {27}\gamma_1\varepsilon^2+\mathcal O(\varepsilon^3)$.

We obtain $\eta = \frac{4}{243}\varepsilon^2$, hence $z=2-\frac{2}{243}\varepsilon^2$ and $2\nu = 1+\tfrac 16 \varepsilon + [\tfrac{1}{36} -\tfrac 29\gamma_2-\tfrac{2}{27}\gamma_1]\varepsilon^2+\mathcal O(\varepsilon^3)$, with $[\tfrac{1}{36} -\tfrac 29\gamma_2-\tfrac{2}{27}\gamma_1]\approx0.0849$.

\section{Evaluation of loop integrals and expansion about $d_c=2$}
We introduce the loop factor
\begin{align}
	N_d &= \frac{2}{(4\pi)^{d/2}}
	\frac 1 {\Gamma\left(\frac d 2\right)}\;,\nonumber\\
	\int\!\frac{d^d p}{(2\pi)^d}
	&= N_d \int_0^{\infty}dp \;p^{d-1}\;,
\end{align}
with $N_2=1/(2\pi)$ in two dimensions.
Further, the propagator is given by $G_0(\omega_q,q)=\frac{{K_R}q^2}{\omega_q^2+{K_R}q^4}$ and we define $F_q = \sqrt{K_R}\, q^2$.

\subsection{Two-point function}
\label{sec_diagramA}
The setting sun diagram is given by
\begin{align}
A_d(\omega_q,q)
&\;\equiv\;
\int\!\frac{d\omega_k}{2\pi}\,\frac{d^d k}{(2\pi)^d}\,
\int\!\frac{d\omega_p}{2\pi}\,\frac{d^d p}{(2\pi)^d}\,
\nonumber\\
&\quad
G_0(\omega_k,k)\,G_0(\omega_p,p)\,
\nonumber\\
&\quad
G_0(\omega_q-\omega_k-\omega_p,q-k-p)\;.
\end{align}
Using the convolution identity (for $F_1,F_2\in\mathbb R^+$)
\begin{align}\label{conv_id}
&\int_{-\infty}^{\infty}\!\frac{d\omega}{2\pi}\;
\frac{1}{(\omega^2+F_1^2)\big[(\omega_x-\omega)^2+F_2^2\big]}
\nonumber\\
&=
\frac{F_1+F_2}{2F_1F_2\big(\omega_x^2+(F_1+F_2)^2\big)},
\end{align}
first with $F_1=F_p$, $F_2=F_{q-p-k}$, $\omega_x=\omega_q-\omega_k$, and then over $\omega_k$ with
$F_1=F_k$, $F_2=F_p+F_{q-p-k}$, $\omega_x=\omega_q$, we obtain
\begin{align}
A_d(\omega_q,q)
&=\frac{{K_R}^{3/2}}{4}
\int\!\frac{d^d p}{(2\pi)^d}\frac{d^d k}{(2\pi)^d}\;
\frac{\mathcal F(q,p,k)}
{\omega_q^2+\mathcal F(q,p,k)^2},
\nonumber\\
\mathcal F(q,p,k)&:=F_p+F_k+F_{q-p-k}.
\end{align}
In the static limit $\omega_q=0$, we can evaluate the double momentum integral as
\begin{align}
A_d(0,q)
&=\frac{{K_R}}{4}\int\!\frac{d^d p}{(2\pi)^d}
\frac{d^d k}{(2\pi)^d}\;
\frac{1}{(q-p-k)^2+p^2+k^2}
\nonumber\\
&= \frac{{K_R}}{16}
3^{\,1-\frac{3d}{2}}\, N_d^2\,
\Gamma(1-d)\,\Gamma\left(\frac d 2 \right)^2
\big(q^2\big)^{d-1}\;.
\end{align}
Now, in order to obtain the field renormalization, we calculate 
\begin{align}
\partial_{q^2}A_d(0,q)
&= - \frac{{K_R}}{16}
3^{\,1-\frac{3d}{2}}\, N_d^2\,
\nonumber\\
&\quad
\Gamma(2-d)\,\Gamma\left(\frac d 2 \right)^2
\big(q^2\big)^{d-2}
\end{align}
which, in $d=2-\varepsilon$ in $\overline{MS}$ scheme (see below), yields
\begin{align}
	\label{diagramA_MS}
\partial_{q^2}A_{2-\varepsilon}^{(\overline{MS})}(0,q)
&=
-\frac{{K_R}\,N_2^{2}}{144}\,\frac{1}{\varepsilon}
\Big[
1
+\varepsilon\Big(\tfrac{3}{2}\ln 3-\ln(q^{2})\Big)
\Big]
\nonumber\\
&\quad+\mathcal{O}(\varepsilon)
\;.
\end{align}

\subsection{Four-point function - first diagram}
\label{sec_diagramB}
The one-loop diagram correction to the four-point interaction reads
\begin{align}
	&B(\omega_q,q)
	\;\equiv\;
	\int\!\frac{d\omega_p}{2\pi}\,\frac{d^d p}{(2\pi)^d}\,
	G_0(\omega_p,p)\,G_0(\omega_p+\omega_q,p+q)
	\nonumber\\
	&\qquad= \frac{K_R}{2}\,\int_p
	\frac{F_p+F_{p+q}}
	{\omega_q^2+ (F_p+F_{p+q})^2}\;.
\end{align}
For $\omega_q=0$, we find
\begin{align}
	B_d(0,q)
	= \frac{\sqrt{K_R}}{2^{d+1}} N_d \Gamma\Big(\frac d 2\Big) \Gamma\Big(1-\frac d 2\Big) (q^2)^{d/2 -1} \;.
\end{align}
Now, we apply the $\overline{MS}$-scheme, i.e., multiply the result by $N_2/N_d$. Hence, in $d=2-\varepsilon$ dimensions, the Laurent series reads to leading order
\begin{align}\label{diagramB_MS}
	B^{\overline{MS}}_{2-\varepsilon}(q) = b_1+B_r(q)+\mathcal O(\varepsilon) \;,
\end{align}
where we separated $b_1$ the divergent part in $\varepsilon$, from the momentum dependent (IR divergent) part $B_r(q)$
\begin{align}\label{diagramB_MS_split}
	b_1=\frac{\sqrt{K_R}}{4} N_2 \frac{1}{\varepsilon}\;,\qquad 
	B_r(q)=-\frac{\sqrt{K_R}}{4} N_2\, [\frac{1}{2}\ln q^2-\ln 2]\;.
\end{align}

\subsection{Four-point function - second diagram}
\label{sec_diagramC}
The last two-loop diagram we need is given by
\begin{widetext}
\begin{align}
	\tilde C_d(\omega_{q_1},q_1,0,0) 
	\;\equiv\;
	\int\!\frac{d\omega_k}{2\pi}\,\frac{d^d k}{(2\pi)^d}\,
	\int\!\frac{d\omega_p}{2\pi}\,\frac{d^d p}{(2\pi)^d}\,
	G_0(\omega_p+\omega_{q_1},p+q_1)\,G_0(\omega_p,p)\,G_0(\omega_p-\omega_k,p-k)\,G_0(\omega_k,k)\;.
\end{align}
\end{widetext}
We integrate out $\omega_k$ making use of the convolution~\eqref{conv_id}, then using
\begin{widetext}
\begin{align}
	&\int_{-\infty}^{\infty}\frac{d\omega_p}{2\pi}\,
	\frac{1}{(\omega_p^2+F_p^2)\big[(\omega_p+\omega_q)^2+F_{p+q}^2\big](\omega_p^2+F_x^2)}\nonumber\\
	&\qquad=\frac{1}{2\,(F_x^{2}-F_p^{2})}
	\!\left[
	\frac{1}{F_p\big((\omega_q+i F_p)^2+F_{p+q}^2\big)}
	-\frac{1}{F_x\big((\omega_q+i F_x)^2+F_{p+q}^2\big)}
	\right]\nonumber\\
	&\qquad\quad+\frac{1}{2 F_{p+q}\,\big((-\omega_q+i F_{p+q})^2+F_p^2\big)\,\big((-\omega_q+i F_{p+q})^2+F_x^2\big)}\,,
\end{align}
\end{widetext}
with additionally $\omega_{q_1}=0$ and $C_d(q)=\tilde C_d(0,q,0,0)$, we find the expression in terms of momentum integrals to be
\begin{widetext}
\begin{align}
	C_d(q)
	&= \frac{K_R^2}{4} \int_{p,k}
	\frac{F_p+F_{p+q}+F_{p-k}+F_k}
	{(F_p+F_{p+q})\,(F_{p+q}+F_{p-k}+F_k)\,(F_p+F_{p-k}+F_k)}\\
	&=\;\frac{K_R}{4}\!\int\!\frac{d^d p}{(2\pi)^d}\frac{d^d k}{(2\pi)^d}\,
	\frac{p^2+(p+q)^2+(p-k)^2+k^2}
	{\big[p^2+(p+q)^2\big]\big[(p+q)^2+(p-k)^2+k^2\big]\big[p^2+(p-k)^2+k^2\big]}\;.
\end{align}
\end{widetext}
Introduce
\begin{align*}
	D_{12}&:=p^{2}+(p+q)^{2},\\
	D_{23}&:=(p+q)^{2}+(p-k)^{2}+k^{2},\\
	D_{13}&:=p^{2}+(p-k)^{2}+k^{2}\;,
\end{align*}
such that the integrand splits as
\begin{align}
	&\frac{p^{2}+(p+q)^{2}+(p-k)^{2}+k^{2}}
	{D_{12}D_{13}D_{23}}\\
	&\quad=\frac12\!\left(
    \frac{1}{D_{13}D_{23}}
	+\frac{1}{D_{12}D_{23}}
	+\frac{1}{D_{12}D_{13}}
	\right).
	\label{eq:split}
\end{align}
With a Feynman parameter, we can perform the $p$ integration and obtain
\begin{align}
	C_{d}(q)&=
	\frac{K_R}{2^5(4\pi)^{d/2}}\,
	\Gamma\!\Big(2-\frac{d}{2}\Big)
	\nonumber\\
	&\quad
	\int \frac{\dd^{d}k}{(2\pi)^{d}}\,
	\sum_{ij\in\{12,13,23\}}\int_{0}^{1}\!\dd x\;
	\nonumber\\
	&\quad
	\big(\widetilde S_{ij}(x)\big)^{\alpha_p},\quad
	\alpha_p:=\frac{d}{2}-2,
	\label{eq:Cq-after-p}
\end{align}
with the shorthand
\begin{align}
	\widetilde S_{ij}(x)
	&=A_{ij}(x)\,Q^{2}+B_{ij}(x)\,r^{2}
	+C_{ij}(x)\,Q\,r\,\mu,\nonumber\\
	Q&:=|\bm q|,\;\;r:=|\bm k|,\nonumber\\
	\mu&:=\frac{\bm q\!\cdot\!\bm k}{Q\,r},
\end{align}
and the channel polynomials 
\begin{align}
	(12):\;& A_{12}=(1-x^2)/4,\quad B_{12}=3/4,\nonumber\\
	& C_{12}=(1-x)/2,\\
	(13):\;& A_{13}=1/4,\quad B_{13}=(3-2x-x^{2})/4,\nonumber\\
	& C_{13}=(1-x)/2,\nonumber\\
	(23):\;& A_{23}=x(2-x)/4,\nonumber\\
	& B_{23}=(3-2x-x^{2})/4,\quad C_{23}=x(1-x)/2\;.\nonumber
\end{align}
\subsubsection{Angular and radial $k$–integrations}
For $\Re a>0$,
\begin{align}
	\int \dd\Omega_{d-1}\,(a+b\mu)^{\alpha}
	&=\frac{2\pi^{d/2}}{\Gamma(d/2)}\,a^{\alpha}\;
	\nonumber\\
	&\quad
	{}_2F_{1}\!\Big(-\tfrac{\alpha}{2},\tfrac{1-\alpha}{2};\tfrac d2;
	\frac{b^{2}}{a^{2}}\Big),
\end{align}
with $a=A_{ij}Q^{2}+B_{ij}r^{2}$, $b=C_{ij}Q\,r$ and using the hypergeometric function $_2F_1$. Hence
\begin{align}
	C_{d}(q)
	&=\frac{K_R\,\Gamma(2-\tfrac d2)}{2^5(4\pi)^{d/2}}\;
	\frac{2\pi^{d/2}}{\Gamma(d/2)}\,
	\frac{1}{(2\pi)^{d}}
	\nonumber\\
	&\quad
	\sum_{ij}\int_{0}^{1}\!\dd x
	\int_{0}^{\infty}\!\dd r\,r^{d-1}\,(A_{ij}Q^{2}+B_{ij}r^{2})^{\alpha_p}
	\nonumber\\[-2pt]
	&\quad\times
	{}_2F_{1}\!\left(-\tfrac{\alpha_p}{2},\tfrac{1-\alpha_p}{2};\tfrac d2;
	\frac{C_{ij}^{2}\,Q^{2}r^{2}}{(A_{ij}Q^{2}+B_{ij}r^{2})^{2}}\right).
	\label{eq:Cq-r}
\end{align}
Let $t=r^{2}$ ($r^{d-1}\dd r=\tfrac12\,t^{\frac d2-1}\dd t$) and map
\begin{align}
	y&=\frac{B_{ij}t}{A_{ij}Q^{2}+B_{ij}t}\in(0,1),
	\nonumber\\
	\frac{C_{ij}^{2}Q^{2}t}{(A_{ij}Q^{2}+B_{ij}t)^{2}}
	&=\frac{C_{ij}^{2}}{A_{ij}B_{ij}}\,y(1-y).
\end{align}
together with $\frac{\dd y}{y(1-y)}=\frac{\dd t}{t}$.
After simplifying constants, one finds
\begin{align}
	C_{d}(q)&=\frac{K_R\,\Gamma(2-\tfrac d2)}
	{2^{\,2d+5}\,\pi^{d}\,\Gamma(\tfrac d2)}\,
	Q^{2(d-2)}
	\nonumber\\
	&\quad
	\sum_{ij}\int_{0}^{1}\!\dd x\;A_{ij}^{\,d-2}\,B_{ij}^{-d/2}\;
	I_{y}\!\Big(\lambda_{ij}(x)\Big),
	\label{eq:Cq-y}
\end{align}
where, for clarity of notation, $\alpha=\tfrac d2$, $\beta=2-\tfrac d2$ and
\begin{align}
	I_{y}(\lambda)
	&=\int_{0}^{1}\! y^{\alpha-1}(1-y)^{\beta-\alpha-1}\;
	\nonumber\\
	&\quad
    {}_2F_{1}\!\big(\tfrac{\beta}{2},\tfrac{\beta+1}{2};\alpha;\ \lambda\,y(1-y)\big)\,\dd y,
    \nonumber\\
	\lambda_{ij}&=\frac{C_{ij}^{2}}{A_{ij}B_{ij}}\;.
\end{align}

\subsubsection{Formulas used}
In the following two sections, we make extensive use of the integral representations and series expansions in terms of Pochhammer symbols of the hypergeometric function $_2F_1$, the Appell series, and (incomplete) Beta-functions. We provide the necessary expressions, consult for instance~\cite{Gradshteyn2015}.

Definition of the Appell F1
\begin{align}\label{eq:AppellF1}
	F_{1}(a;\,b_{1},b_{2};\,c;\,x,y)
	&=\frac{\Gamma(c)}{\Gamma(a)\,\Gamma(c-a)}
	\nonumber\\
	&\quad
	\int_{0}^{1}
	u^{\,a-1}\,(1-u)^{\,c-a-1}\,
	\nonumber\\
	&\quad
	(1-xu)^{-b_{1}}\,(1-yu)^{-b_{2}}\;\mathrm{d}u,
\end{align}
valid for $ \Re(c)>\Re(a)>0$. The series expansion is
\begin{align}
	F_{1}(a;\,b_{1},b_{2};\,c;\,x,y)
	&=\sum_{m=0}^{\infty}\sum_{n=0}^{\infty}
	\frac{(a)_{m+n}\,(b_{1})_{m}\,(b_{2})_{n}}{(c)_{m+n}}\,
	\nonumber\\
	&\quad
	\frac{x^{m}}{m!}\,\frac{y^{n}}{n!},\quad (|x|<1,\ |y|<1)\;,
\end{align}
given in terms of the rising Pochhammer symbols
\begin{align}
	(q)_{0}&=1,\nonumber\\
	(q)_{n}&=q(q+1)\cdots(q+n-1)
	=\frac{\Gamma(q+n)}{\Gamma(q)}.
\end{align}
The incomplete Beta-function is given by
\begin{align}
	B_z(a,b)
	&= \int_{0}^{z} t^{a-1}\,(1-t)^{b-1}\,\dd t
	\nonumber\\
	&= \frac{z^{a}}{a}\;
	{}_2F_{1}\!\left(a,\,1-b;\,a+1;\,z\right).
\end{align}
And for $z=1$, the Beta-function can be given in terms of Gamma-functions
\begin{align}
	B(a,b)
	&= \int_{0}^{1} t^{a-1}\,(1-t)^{b-1}\,\dd t
	= \frac{\Gamma(a)\,\Gamma(b)}{\Gamma(a+b)}\;.
\end{align}

\subsubsection{$y$–integration (closed form)}
Expanding the Gauss function and integrating term wise, we obtain
\begin{align}
	\int_0^1 y^{\alpha-1+n}&(1-y)^{\beta-\alpha-1+n}\dd y
	=B(\alpha+n,\beta-\alpha+n)
	\nonumber\\
	&\qquad= \frac{\Gamma(\alpha)\Gamma(\beta-\alpha)}{\Gamma(\beta)}
	\frac{(\alpha)_n(\beta-\alpha)_n}{(\beta)_{2n}}\;.
\end{align}
Using $(\beta)_{2n}=2^{2n}\,(\tfrac{\beta}{2})_n(\tfrac{\beta+1}{2})_n$, the Pochhammer symbols cancel and the series sums to a binomial
\begin{align}
	   \sum_{n=0}^\infty \frac{(\tfrac{\beta}{2})_n\,(\tfrac{\beta+1}{2})_n}{(\alpha)_n\,n!}\,B(\alpha+n,\beta-\alpha+n)\,\lambda^n\\
   =\frac{\Gamma(\alpha)\Gamma(\beta-\alpha)}{\Gamma(\beta)}\Big(1-\frac{\lambda}{4}\Big)^{\alpha-\beta}\;.
\end{align}
Inserting into~\eqref{eq:Cq-y} and canceling $\Gamma(\tfrac d2)$ gives the prefactor
\begin{align}
	C_{d}(q)&=\frac{K_R\,\Gamma\!\big(2-d\big)}{2^{\,2d+5}\,\pi^{d}}\,
	Q^{2(d-2)}\nonumber\\
	&\qquad\sum_{ij}\int_{0}^{1}\!\dd x\;A_{ij}^{\,d-2}\,B_{ij}^{-d/2}\;
	\Big(1-\frac{\lambda_{ij}(x)}{4}\Big)^{d-2}.
	\label{eq:Cq-3F2-correct}
\end{align}

\subsubsection{Series expansion and $x$–integration: final double sum}
Using
\begin{equation}
	A_{ij}^{\,d-2}B_{ij}^{-d/2}\Big(\tfrac{\lambda_{ij}}{4}\Big)^{n}
	=\frac{C_{ij}^{2n}}{4^{\,n}}\;A_{ij}^{\,d-2-n}\,B_{ij}^{-d/2-n},
\end{equation}
define the channel moments
\begin{equation}
	\mathcal I^{(ij)}_{n}:=\int_{0}^{1}\!\dd x\;
	C_{ij}(x)^{2n}\,A_{ij}(x)^{\,d-2-n}\,B_{ij}(x)^{-d/2-n}.
	\label{eq:In-def}
\end{equation}
Then \eqref{eq:Cq-3F2-correct} becomes the explicit double sum
\begin{align}
	C_{d}(q)&=\frac{K_R\,\Gamma(2-d)}{2^{\,2d+5}\,\pi^{d}}\,
	Q^{2(d-2)}\nonumber\\
	&\qquad\sum_{n=0}^{\infty}\frac{(2-d)_n}{n!\,4^{\,n}}
	\;\sum_{ij\in\{12,13,23\}}\mathcal I^{(ij)}_{n}.
	\label{eq:Cq-double-sum}
\end{align}
we can evaluate the different channels by means of substitutions, which we give for each function and representations of corresponding hypergeometric functions.
For the first channel, we substitute $u=1-x$, such that $1+x=2-u$, and use
$\int_0^1 u^{a-1}(1-zu)^{-b}\,\dd u=\tfrac1a\,{}_2F_1(a,b;a+1;z)$ with $z=\tfrac12$, to arrive at\begin{align}
	\mathcal I^{(12)}_{n} &=
	\frac{2^{n+2}\;3^{\,-\tfrac d2 -n}}{\,d-1+n\,}\;
	\nonumber\\
	&\quad
	{}_2F_{1}\!\left(d-1+n,\;2-d+n;\;d+n;\;\tfrac12\right)\;,
\end{align}
in the second, again with $x=1-2u$, then $u=2v$, to obtain
\begin{align}
	\mathcal I^{(13)}_{n} &=
	\frac{2^{-\,2d+4}}{\,n+1-\tfrac d2\,}\;
	\nonumber\\
	&\quad
	{}_2F_{1}\!\left(n+1-\tfrac d2,\;n+\tfrac d2;\;n+2-\tfrac d2;\;\tfrac14\right)\;,
\end{align}
and finally, in the third, no simplification could be found, such that we arrive at
\begin{widetext}
\begin{align}
	\mathcal I^{(23)}_{n} =
	2^{\,n+2}\,3^{-\,\frac d2 - n}\;
	\frac{\Gamma(n+d-1)\,\Gamma\!\big(n+1-\tfrac d2\big)}{\Gamma\!\big(2n+\tfrac d2\big)}\;
	F_{1}\!\Big(n+d-1;\,n+2-d,\,\tfrac d2+n;\,2n+\tfrac d2;\,\tfrac12,\,-\tfrac13\Big).
\end{align}
\end{widetext}
with $F_1$ being the Appell F1 function, compare Equation~\eqref{eq:AppellF1}.

\subsubsection{Expansion in \(\overline{MS}\) scheme}
As before, we multiply with $(N_2/N_d)^2$ and obtain
\begin{align}
	C_{d}^{(\overline{MS})}(q)&=\frac{K_R\,N_2^2\,\Gamma(2-d)\Gamma(\tfrac d2)^2}{2^{7}}\,
	Q^{2(d-2)}\nonumber\\
	&\qquad\sum_{n=0}^{\infty}\frac{(2-d)_n}{n!\,4^{\,n}}
	\;\sum_{ij\in\{12,13,23\}}\mathcal I^{(ij)}_{n}.
	\label{eq:Cq-double-sum-MS}
\end{align}
In order to expand the expression \eqref{eq:Cq-double-sum-MS}, we represented the channels in terms of (double) sums and expanded them to the necessary order in $\varepsilon$, we will find that
\begin{align}
	\mathcal I_0^{(12)} &= \frac{4}{3}+\mathcal O(\varepsilon)\;,\\	
	\mathcal I_0^{(13)} &= \frac{2}{\varepsilon}+6\ln 2-\ln 3
	+\mathcal O(\varepsilon)\nonumber\\
	&= \mathcal I_0^{(23)}\;,\nonumber\\
	\mathcal I_n^{(ij)} &= \mathcal O(\varepsilon^0)\;, \forall n>0,ij\in\{12,13,23\}\;.
\end{align}
We used several times Equation~\eqref{eq:trick}, and arrive at
\begin{align}\label{diagramC_MS}
	C_{2-\varepsilon}^{(\overline{MS})}(q)
	&=\frac{K_R\,N_2^{2}}{2^{7}}
	\Big[\frac 4 {\varepsilon^{2}}+(\tfrac{4}{3}+12\ln 2-2\ln 3
	- 4\ln q^2)\!\frac{1}{\varepsilon}\nonumber\\
	&\qquad\quad
	+\mathcal O(\varepsilon^0)
	\Big]\;.
\end{align}

\subsubsection{Used trick on expansion for the hypergeometric functions}
Let $a=\mathcal O(\varepsilon)$, and $b=b_0+\mathcal O(\varepsilon)=c$, where $b_0$ is a non-zero constant and $|z|<1$, we want to show that the hypergeometric series,
\begin{equation}
	{}_2F_{1}(a,b;c;z)
	= \sum_{n=0}^{\infty} \frac{(a)_n\,(b)_n}{(c)_n}\,\frac{z^n}{n!}\,,
\end{equation}
gives
\begin{equation}\label{eq:trick}
	{}_2F_{1}(a,b;c;z)
	= 1 - a \ln(1-z) + \mathcal{O}(\varepsilon^2)\,.
\end{equation}
to linear order in $\varepsilon$.

Note, for $n=0$ the term is $1$ and has no $\varepsilon$--dependence. For every $n\ge1$, $(a)_n$ is $\mathcal{O}(a)=\mathcal{O}(\varepsilon)$, while the deviations of $b$ and $c$ from $b_0$ are also $\mathcal{O}(\varepsilon)$. Thus, any correction from $b$ or $c$ would multiply $a$ and be $\mathcal{O}(\varepsilon^2)$, which we can drop at linear order.

Therefore, to $\mathcal{O}(\varepsilon)$ we may set $b=c=b_0$ in the $n\ge1$ terms. Further, for small $a$, one has $(a)_n = a(a+1)\cdots(a+n-1) = a\,(n-1)! + \mathcal{O}(a^2)$.
Hence
\begin{align}
	\sum_{n=1}^{\infty}\frac{(a)_n}{n!}\,z^n
	&= a\sum_{n=1}^{\infty}\frac{z^n}{n} + \mathcal{O}(a^2)
	\nonumber\\
	&
	= a\big(-\ln(1-z)\big) + \mathcal{O}(a^2)\,,
\end{align}
and therefore the claim is shown.

\bibliography{main}
\end{document}